\documentclass[dvipdfmx]{preprint}
\usepackage{geometry}
\usepackage{color}
\definecolor{note_fontcolor}{rgb}{0.578125, 0.542969, 0.542969}
\definecolor{shadecolor}{rgb}{0.941406, 0.941406, 0.847656}
\usepackage{verbatim}
\usepackage{framed}
\usepackage{amsmath}
\usepackage{amsthm}
\usepackage{amssymb}
\usepackage{stmaryrd}

\makeatletter

\newenvironment{lyxgreyedout}
  {\textcolor{note_fontcolor}\bgroup\ignorespaces}
  {\ignorespacesafterend\egroup}

\numberwithin{equation}{section}
\theoremstyle{plain}
\newtheorem{thm}{\protect\theoremname}[section]
\theoremstyle{remark}
\newtheorem{rem}[thm]{\protect\remarkname}
\theoremstyle{plain}
\newtheorem{prop}[thm]{\protect\propositionname}
\theoremstyle{plain}
\newtheorem{conjecture}[thm]{\protect\conjecturename}
\theoremstyle{plain}
\newtheorem{lem}[thm]{\protect\lemmaname}
\theoremstyle{definition}
\newtheorem{defn}[thm]{\protect\definitionname}
\theoremstyle{plain}
\newtheorem{cor}[thm]{\protect\corollaryname}
\theoremstyle{definition}
\newtheorem{example}[thm]{\protect\examplename}
\theoremstyle{remark}
\newtheorem*{acknowledgement*}{\protect\acknowledgementname}

\usepackage{fancybox}
\usepackage{textcomp}
\usepackage[utf8]{inputenc}
\usepackage{lmodern}

\usepackage[destlabel,backref,dvipdfmx,bookmarksopenlevel=4,bookmarksnumbered]{hyperref}
\usepackage{color}
\usepackage{bbm}
\usepackage{url}
\definecolor{magenta}{RGB}{30, 0, 50}
\definecolor{shadecolor}{rgb}{0.95,1,0.9}
\DeclareFontFamily{U}{mathx}{\hyphenchar\font45}
\DeclareFontShape{U}{mathx}{m}{n}{
      <5> <6> <7> <8> <9> <10>
      <10.95> <12> <14.4> <17.28> <20.74> <24.88>
      mathx10
      }{}
\DeclareSymbolFont{mathx}{U}{mathx}{m}{n}

\DeclareMathSymbol{\bigtimes}       {1}{mathx}{"91}%"



\DeclareMathOperator{\indef}{indef}

\makeatother

\providecommand{\acknowledgementname}{Acknowledgement}
\providecommand{\conjecturename}{Conjecture}
\providecommand{\corollaryname}{Corollary}
\providecommand{\definitionname}{Definition}
\providecommand{\examplename}{Example}
\providecommand{\lemmaname}{Lemma}
\providecommand{\propositionname}{Proposition}
\providecommand{\remarkname}{Remark}
\providecommand{\theoremname}{Theorem}

\begin{document}
\title{Antinormally-Ordered Quantizations, phase space path integrals and
the Olshanski semigroup of a symplectic group.}
\author{Hideyasu Yamashita}
\institute{Division of Liberal Arts and Sciences, Aichi-Gakuin University\\
\email{yamasita@dpc.aichi-gakuin.ac.jp}}

\maketitle

%% disable shaded env
\renewenvironment{shaded}
  {\bgroup\ignorespaces}
  {\ignorespacesafterend\egroup}

\newenvironment{trivenv}
  {\bgroup\ignorespaces}
  {\ignorespacesafterend\egroup}

\newcommand{\displabel}[1]{\textcolor[rgb]{0.5, 0.5, 0.5}{\texttt{\textup{\tiny{}lab=#1}}}}

\newcommand{\hidable}[3]{#2}
\newcommand{\hidea}[1]{{#1}}
\newcommand{\hideb}[1]{{#1}}
\newcommand{\hidec}[1]{{#1}}
\newcommand{\hidep}[1]{{#1}}
\renewcommand{\hidec}[1]{}
\renewcommand{\hidep}[1]{}

\newcommand{\thlab}[1]{{\tt [#1]}}

\newenvironment{proofbar}
{\begin{leftbar}\noindent{\bf Proof.}}
{\noindent{\bf QED}\end{leftbar}}

\global\long\def\N{\mathbb{N}}%
\global\long\def\C{\mathbb{C}}%
\global\long\def\Z{\mathbb{Z}}%
 
\global\long\def\R{\mathbb{R}}%
 
\global\long\def\im{\mathrm{i}}%

\global\long\def\di{\partial}%
 
\global\long\def\d{{\rm d}}%

\global\long\def\ol#1{\overline{#1}}%
\global\long\def\ul#1{\underline{#1}}%
\global\long\def\ob#1{\overbrace{#1}}%

\global\long\def\ov#1{\overline{#1}}%

\global\long\def\then{\Rightarrow}%
 
\global\long\def\Then{\Longrightarrow}%

\global\long\def\al{\alpha}%
\global\long\def\de{\delta}%
 
\global\long\def\ep{\epsilon}%
 
\global\long\def\la{\lambda}%
 
\global\long\def\io{\iota}%
 
\global\long\def\th{\theta}%
\global\long\def\si{\sigma}%
 
\global\long\def\om{\omega}%

\global\long\def\De{\Delta}%
 
\global\long\def\Th{\Theta}%
 
\global\long\def\Om{\Omega}%

\global\long\def\brho{\boldsymbol{\rho}}%
\global\long\def\bDelta{\boldsymbol{\Delta}}%
 
\global\long\def\bmu{\boldsymbol{\mu}}%
 
\global\long\def\bchi{\boldsymbol{\chi}}%
 
\global\long\def\bPi{\boldsymbol{\Pi}}%
 
\global\long\def\bOm{\boldsymbol{\Omega}}%

\global\long\def\cA{\mathcal{A}}%
\global\long\def\cB{\mathcal{B}}%
 
\global\long\def\cC{\mathcal{C}}%
 
\global\long\def\cD{\mathcal{D}}%
\global\long\def\cE{\mathcal{E}}%
 
\global\long\def\cF{\mathcal{F}}%
 
\global\long\def\cG{{\cal G}}%
 
\global\long\def\cH{\mathcal{H}}%
 
\global\long\def\cI{\mathcal{I}}%
 
\global\long\def\cJ{\mathcal{J}}%
\global\long\def\cK{\mathcal{K}}%
 
\global\long\def\cL{\mathcal{L}}%
 
\global\long\def\cM{\mathcal{M}}%
 
\global\long\def\cN{\mathcal{N}}%
 
\global\long\def\cO{\mathcal{O}}%
 
\global\long\def\cP{\mathcal{P}}%
 
\global\long\def\cQ{\mathcal{Q}}%
 
\global\long\def\cR{\mathcal{R}}%
 
\global\long\def\cS{\mathcal{S}}%
 
\global\long\def\cT{\mathcal{T}}%
 
\global\long\def\cU{\mathcal{U}}%
 
\global\long\def\cV{\mathcal{V}}%
 
\global\long\def\cW{\mathcal{W}}%
\global\long\def\cX{\mathcal{X}}%
 
\global\long\def\cY{\mathcal{Y}}%
 
\global\long\def\cZ{\mathcal{Z}}%

\global\long\def\scA{\mathscr{A}}%
\global\long\def\scB{\mathscr{B}}%
\global\long\def\scC{\mathscr{C}}%
\global\long\def\scD{\mathscr{D}}%
 
\global\long\def\scE{\mathscr{E}}%
 
\global\long\def\scF{\mathscr{F}}%
 
\global\long\def\scG{\mathscr{G}}%
 
\global\long\def\scH{\mathscr{H}}%
 
\global\long\def\scI{\mathscr{I}}%
 
\global\long\def\scJ{\mathscr{J}}%
 
\global\long\def\scK{\mathscr{K}}%
 
\global\long\def\scL{\mathscr{L}}%
 
\global\long\def\scM{\mathscr{M}}%
 
\global\long\def\scN{\mathscr{N}}%
 
\global\long\def\scO{\mathscr{O}}%
 
\global\long\def\scP{\mathscr{P}}%
 
\global\long\def\scR{\mathscr{R}}%
\global\long\def\scS{\mathscr{S}}%
 
\global\long\def\scT{\mathscr{T}}%
 
\global\long\def\scU{\mathscr{U}}%
 
\global\long\def\scZ{\mathscr{Z}}%

\global\long\def\bbA{\mathbb{A}}%
 
\global\long\def\bbB{\mathbb{B}}%
 
\global\long\def\bbD{\mathbb{D}}%
 
\global\long\def\bbF{\mathbb{F}}%
 
\global\long\def\bbG{\mathbb{G}}%
 
\global\long\def\bbI{\mathbb{I}}%
 
\global\long\def\bbK{\mathbb{K}}%
 
\global\long\def\bbL{\mathbb{L}}%
 
\global\long\def\bbM{\mathbb{M}}%
 
\global\long\def\bbP{\mathbb{P}}%
 
\global\long\def\bbQ{\mathbb{Q}}%
 
\global\long\def\bbT{\mathbb{T}}%
 
\global\long\def\bbU{\mathbb{U}}%
 
\global\long\def\bbX{\mathbb{X}}%
 
\global\long\def\bbY{\mathbb{Y}}%
\global\long\def\bbW{\mathbb{W}}%

\global\long\def\bbOne{1\kern-0.7ex  1}%
 %defined as 1\kern-0.7ex1

\renewcommand{\bbOne}{\mathbbm{1}}

\global\long\def\bB{\mathbf{B}}%
 
\global\long\def\bG{\mathbf{G}}%
 
\global\long\def\bH{\mathbf{H}}%
 
\global\long\def\bM{\mathbf{M}}%
 
\global\long\def\bS{\boldsymbol{S}}%
 
\global\long\def\bT{\mathbf{T}}%
 
\global\long\def\bX{\mathbf{X}}%
\global\long\def\bY{\mathbf{Y}}%
\global\long\def\bW{\mathbf{W}}%
 
\global\long\def\boT{\boldsymbol{T}}%

\global\long\def\fraka{\mathfrak{a}}%
 
\global\long\def\frakb{\mathfrak{b}}%
 
\global\long\def\frakc{\mathfrak{c}}%
 
\global\long\def\frake{\mathfrak{e}}%
 
\global\long\def\frakf{\mathfrak{f}}%
 
\global\long\def\fg{\mathfrak{g}}%
 
\global\long\def\frakh{\mathfrak{h}}%
 
\global\long\def\fraki{\mathfrak{i}}%
 
\global\long\def\frakk{\mathfrak{k}}%
 
\global\long\def\frakl{\mathfrak{l}}%
 
\global\long\def\frakm{\mathfrak{m}}%
 
\global\long\def\frakn{\mathfrak{n}}%
 
\global\long\def\frako{\mathfrak{o}}%
 
\global\long\def\frakp{\mathfrak{p}}%
 
\global\long\def\frakq{\mathfrak{q}}%
 
\global\long\def\frakr{\mathfrak{r}}%
 
\global\long\def\fs{\mathfrak{s}}%
 
\global\long\def\frakt{\mathfrak{t}}%
 
\global\long\def\fraku{\mathfrak{u}}%

\global\long\def\fA{\mathfrak{A}}%
 
\global\long\def\fB{\mathfrak{B}}%
 
\global\long\def\fC{\mathfrak{C}}%
 
\global\long\def\fD{\mathfrak{D}}%
 
\global\long\def\fF{\mathfrak{F}}%
 
\global\long\def\fG{\mathfrak{G}}%
 
\global\long\def\fK{\mathfrak{K}}%
 
\global\long\def\fL{\mathfrak{L}}%
 
\global\long\def\fM{\mathfrak{M}}%
 
\global\long\def\fP{\mathfrak{P}}%
 
\global\long\def\fR{\mathfrak{R}}%
 
\global\long\def\fT{\mathfrak{T}}%
 
\global\long\def\fU{\mathfrak{U}}%
 
\global\long\def\fX{\mathfrak{X}}%

\global\long\def\hM{\hat{M}}%

\global\long\def\rM{\mathrm{M}}%
\global\long\def\prj{\mathfrak{P}}%

{} 
\global\long\def\sy#1{{\color{blue}#1}}%

\global\long\def\magenta#1{{\color{magenta}#1}}%

% \global\long\def\symb#1{{\color{red}#1}}%
\global\long\def\symb#1{#1}%

{} %

\global\long\def\emhrb#1{\text{{\color{red}\huge{\bf #1}}}}%

\newcommand{\symbi}[1]{\index{$ #1$}{\color{red}#1}} 

{} 
% \global\long\def\SYM#1#2{\symb{#1}_{\##2}}%
\global\long\def\SYM#1#2{#1}%

\renewcommand{\SYM}[2]{\symb{#1}}

\newcommand{\usuji}{\color[rgb]{0.7,0.4,0.4}} \newcommand{\usu}{\color[rgb]{0.5,0.2,0.1}}
\newenvironment{Usuji} {\begin{trivlist}   \item \usuji }  {\end{trivlist}}
\newenvironment{Usu} {\begin{trivlist}   \item \usu }  {\end{trivlist}} 

\newcommand{\term}[1]{\textcolor[rgb]{0, 0, 1}{\bf #1}}
\newcommand{\termi}[1]{{\bf #1}}

\global\long\def\rG{\mathrm{G}}%
 
\global\long\def\rT{\mathrm{T}}%
 
\global\long\def\rH{\mathrm{H}}%
 
\global\long\def\rU{\mathrm{U}}%

\global\long\def\supp{{\rm supp}}%
\global\long\def\dom{\mathrm{dom}}%
\global\long\def\ran{\mathrm{ran}}%
 
\global\long\def\leng{\text{{\rm leng}}}%
 
\global\long\def\diam{\text{{\rm diam}}}%
 
\global\long\def\Leb{\text{{\rm Leb}}}%
 
\global\long\def\meas{\text{{\rm meas}}}%
\global\long\def\sgn{{\rm sgn}}%
 
\global\long\def\Tr{{\rm Tr}}%
 
\global\long\def\tr{\mathrm{tr}}%
 
\global\long\def\spec{{\rm spec}}%
 
\global\long\def\Ker{{\rm Ker}}%
 
\global\long\def\Lip{{\rm Lip}}%
 
\global\long\def\Id{{\rm Id}}%
 
\global\long\def\id{{\rm id}}%

\global\long\def\ex{{\rm ex}}%
 
\global\long\def\Pow{\mathsf{P}}%
 
\global\long\def\Hom{\mathrm{Hom}}%
 
\global\long\def\grad{\mathrm{grad}}%
 
\global\long\def\End{{\rm End}}%
 
\global\long\def\Aut{{\rm Aut}}%

\newcommand{\slim}{\mathop{\mbox{\rm s-lim}}} %

\newcommand{\wlim}{\mathop{\mbox{\rm w-lim}}}

\newcommand{\limsub}{\mathop{\mbox{\rm lim-sub}}}

\global\long\def\bboxplus{\boxplus}%

\renewcommand{\bboxplus}{\mathop{\raisebox{-0.8ex}{\text{\begin{trivenv}\LARGE{}$\boxplus$\end{trivenv}}}}}

\global\long\def\shuff{\sqcup\kern-0.3ex  \sqcup}%

\renewcommand{\shuff}{\shuffle}

\global\long\def\upha{\upharpoonright}%

\global\long\def\ket#1{|#1\rangle}%
 
\global\long\def\bra#1{\langle#1|}%

{} 
\global\long\def\lll{\vert\kern-0.25ex  \vert\kern-0.25ex  \vert}%
 \renewcommand{\lll}{{\vert\kern-0.25ex  \vert\kern-0.25ex  \vert}}

\global\long\def\biglll{\big\vert\kern-0.25ex  \big\vert\kern-0.25ex  \big\vert\kern-0.25ex  }%
 
\global\long\def\Biglll{\Big\vert\kern-0.25ex  \Big\vert\kern-0.25ex  \Big\vert}%

\newcommand{\iiia}[1]{{\left\vert\kern-0.25ex\left\vert\kern-0.25ex\left\vert #1
  \right\vert\kern-0.25ex\right\vert\kern-0.25ex\right\vert}}

\global\long\def\iii#1{\iiia{#1}}%

\global\long\def\Upa{\Uparrow}%
 
\global\long\def\Nor{\Uparrow}%

\newcommand{\vertt}{\kern-0.6ex\vert}
\renewcommand{\Nor}{[\kern-0.16ex ]}

\global\long\def\Prob{\mathbb{P}}%
\global\long\def\Var{\mathrm{Var}}%
\global\long\def\Cov{\mathrm{Cov}}%
\global\long\def\Ex{\mathbb{E}}%
{} %\newcommand{\F}{\mathbf{F}}
\global\long\def\Ae{{\rm a.e.}}%
 
\global\long\def\samples{\bOm}%

\global\long\def\var{\textrm{{\rm var}}}%
\global\long\def\Hol{\text{{\rm Höl}}}%
 
\global\long\def\hvar{\textrm{{\rm -var}}}%
\global\long\def\hHol{\text{{\rm -Höl}}}%

\global\long\def\pvar{p\textrm{{\rm -var}}}%
\global\long\def\pHol{1/p\text{{\rm -Höl}}}%
\global\long\def\frakt{\mathfrak{t}}%

\global\long\def\var{\textrm{{\rm var}}}%
\global\long\def\Hol{\text{{\rm Höl}}}%
 
\global\long\def\hvar{\textrm{{\rm -var}}}%
\global\long\def\hHol{\text{{\rm -Höl}}}%

\global\long\def\rpvar{\mathfrak{p}}%
 
\global\long\def\rpHol{\mathfrak{h}}%

\global\long\def\bOne{{\bf 1}}%

\global\long\def\Disk{\mathbb{D}^{2}}%
\global\long\def\hcG{\hat{\mathcal{G}}}%
\global\long\def\sfC{\mathsf{C}}%

{} 
\global\long\def\crv{\mathfrak{c}}%
\global\long\def\Crv{\mathfrak{C}}%
 
\global\long\def\gE{\mathrm{e}}%
 
\global\long\def\Rot{{\rm Rot}}%

\global\long\def\bbm#1{\mathbbm{#1}}%

\global\long\def\Mat{{\rm Mat}}%

\global\long\def\cbo{{\bf c}}%
 
\global\long\def\reg{{\rm reg}}%

\global\long\def\decoFor{\mathsf{DF}}%
 
\global\long\def\DF{\mathsf{DF}}%

{} 
\global\long\def\modsp{\scT}%
\global\long\def\regStr{\boldsymbol{T}}%

\global\long\def\smoothfuncs{\scC}%
 
\global\long\def\jj{\mathtt{j}}%
 
\global\long\def\scriptj{\mathtt{j}}%

\global\long\def\newNode{\circledast}%
 
\global\long\def\scriptf{\mathtt{f}}%
 
\global\long\def\scripth{\mathtt{h}}%

\global\long\def\p{\mathbf{p}}%
 
\global\long\def\q{\mathbf{q}}%
 
\global\long\def\bA{\mathbf{A}}%
 
\global\long\def\x{\mathbf{x}}%

\global\long\def\div{\mathrm{div}}%
 
\global\long\def\be{\beta}%
 
\global\long\def\La{\Lambda}%
 
\global\long\def\Ga{\Gamma}%

\global\long\def\wick#1{:\!#1\!:}%
 
\global\long\def\dag{\dagger}%

\global\long\def\braket#1{\langle#1\rangle}%
 
\global\long\def\ka{\kappa}%
 
\global\long\def\z{\mathbf{z}}%

\global\long\def\tI{t_{\mathrm{I}}}%
 
\global\long\def\tF{t_{\mathrm{F}}}%

\global\long\def\ActionIntegral{\mathrm{AI}}%

\global\long\def\Hc{h}%

\global\long\def\coherents{\mathbf{c}}%
 
\global\long\def\ssW{\mathsf{W}}%

\global\long\def\Ten{\bullet}%
{} %

\global\long\def\TT{\intercal}%
 \renewcommand{\TT}{\mathsf{T}}

\global\long\def\trit{\vartriangle\!\! t}%

\global\long\def\Killing{{\rm \boldsymbol{\kappa}}}%
 
\global\long\def\spec{{\rm spec}}%

\global\long\def\nnn{\mathfrak{k}}%

\global\long\def\Ad{{\rm Ad}}%

\global\long\def\Gxz{G\cdot x_{0}}%
 
\global\long\def\lbundle{\scL_{\lambda}}%
 
\global\long\def\Hilb{\cH_{\lambda}}%
 
\global\long\def\Image{{\rm Im}}%
 
\global\long\def\tautlog{{\rm taut}}%
 
\global\long\def\sphere{\mathbb{S}}%
 
\global\long\def\Proj{\mathbf{pj}}%
 
\global\long\def\telem{\mathbf{t}}%
 
\global\long\def\sectio{\mathbf{sect}}%

\global\long\def\hwvec{\mathbf{v}_{\lambda}}%
{} 
\global\long\def\Hwvproj{{\bf E}_{\lambda}}%

\global\long\def\lwvec{{\bf w}_{\lambda}}%

\global\long\def\lbundle{\mathscr{L}_{\lambda}}%
 
\global\long\def\Gxz{G\cdot x_{0}}%

\global\long\def\Rtrans{{\rm Rt}}%
 
\global\long\def\Ltrans{{\rm Lt}}%

\global\long\def\Rght{{\rm R}}%
 
\global\long\def\Lft{{\rm L}}%
 
\global\long\def\Casi{{\bf c}}%
 
\global\long\def\Rtrans{\mathscr{T_{\Rght}}}%
 
\global\long\def\Ltrans{\mathscr{T_{{\rm \Lft}}}}%
 
\global\long\def\Maurer{\mathscr{M}}%
 
\global\long\def\du{\underline{{\rm d}}}%
 
\global\long\def\bphi{\boldsymbol{\varphi}}%

{} 
\global\long\def\ss{{\bf r}}%
 
\global\long\def\infspec{{\bf c}_{\lambda}}%

\global\long\def\Roots{\boldsymbol{R}}%
 
\global\long\def\Sn{{\rm sig}}%
 
\global\long\def\lift{{\rm lift}}%
 
\global\long\def\bpi{\boldsymbol{\pi}}%

\global\long\def\Tensor{\boldsymbol{T}}%
 
\global\long\def\bOmega{\boldsymbol{\Omega}}%
 
\global\long\def\VECSP{\mathbb{V}}%
 
\global\long\def\projection{{\rm pr}}%
 
\global\long\def\dissect{\cD}%
 
\global\long\def\bUpsilin{\boldsymbol{\Upsilon}}%
 
\global\long\def\GEOMR{\mathbf{GR}}%

\global\long\def\fextend#1{\hat{#1}}%
 
\global\long\def\eigenval{\varepsilon}%

\global\long\def\vecalt{\check{\chi}}%
 
\global\long\def\Tanalt{\check{\cA}}%
 
\global\long\def\taut{{\rm taut}}%
 
\global\long\def\tautsection{^{!}\Gamma_{{\rm taut}}^{\infty}}%
\global\long\def\dtautsection{^{!}\Gamma_{{\rm taut}^{*}}^{\infty}}%
 
\global\long\def\holodtautsection{^{!}\Gamma_{{\rm taut}^{*}}^{{\rm hol}}}%
  
\global\long\def\Proj{{\bf pr}}%

\global\long\def\dequantize{\cR}%
 
\global\long\def\FCP{{\rm Coh}}%
 
\global\long\def\Manifold{\grave{{\bf M}}}%
\renewcommand{\Manifold}{{\bf M}}

\global\long\def\Ran{{\rm Ran}}%
 
\global\long\def\qproj{\mathsf{P}}%

\global\long\def\Hc{\grave{H}_{{\rm c}}}%

\global\long\def\dequantize{\cR}%
 
\global\long\def\FCP{{\rm Coh}}%
\global\long\def\Coh{{\rm Coh}}%

\global\long\def\Manifold{{\bf M}}%

\global\long\def\EP{E_{\bbP}}%

\global\long\def\Ran{{\rm Ran}}%
 
\global\long\def\qproj{\mathsf{P}}%

\global\long\def\Taut{{\rm Taut}}%
\global\long\def\bbS{\mathbb{S}}%
\global\long\def\vol{{\rm vol}}%

\global\long\def\Kpot{{\bf h}}%

\global\long\def\ela{e_{\io}}%

\global\long\def\INDEX{\cI}%

\global\long\def\CC{\Gamma}%

\global\long\def\INT{\scI}%

\global\long\def\scV{\mathscr{V}}%
 
\global\long\def\loc{{\rm loc}}%

\global\long\def\path{{\bf x}}%
\global\long\def\Path{{\bf X}}%

\global\long\def\no{\boldsymbol{\tau}}%

\global\long\def\dil{{\d^{\prime}}}%
\global\long\def\dill{{\d^{\prime\prime}}}%
\global\long\def\forms{{\bf A}}%
\global\long\def\ptrans{//}%

\global\long\def\Harmonic{{\rm Hm}}%

\def\mininabla{\text{\tiny $\nabla$}}
\def\Denabla{{\Delta_{\mininabla}}}

\global\long\def\foreignlanguage#1#2{#2}

\newcommand{\HS}{{\bf HS}}

\global\long\def\inv{-1}%
\global\long\def\ppa{{\bf t}}%

\global\long\def\GL{{\rm GL}}%

\global\long\def\rU{{\rm U}}%

\global\long\def\Sp{{\rm Sp}}%
\global\long\def\sp{\mathfrak{sp}}%

\global\long\def\Mp{{\rm Mp}}%

\global\long\def\Spn#1{\textup{Sp}(2n,#1)}%
 
\global\long\def\spn#1{\mathfrak{sp}(2n,#1)}%
 
\global\long\def\GSpn{\Gamma\textup{Sp}(2n)}%
 
\global\long\def\osgn{\cO_{2n}}%

\global\long\def\Spcn#1{\textup{Sp}_{c}(2n,#1)}%
 
\global\long\def\spcN#1#2{\mathfrak{sp}_{c}(2#1,#2)}%
 
\global\long\def\spcn#1{\mathfrak{sp}_{c}(2n,#1)}%
 
\global\long\def\GSpcn{\Gamma\textup{Sp}_{c}(2n)}%

\global\long\def\Mpn#1{\textup{Mp}(2n,#1)}%

\global\long\def\nnn{\nu}%
\global\long\def\Bdd{\cB}%

\global\long\def\Diss{\textup{Diss}}%
\global\long\def\SDiss{\textup{SDiss}}%
\global\long\def\Pos{\textup{Pos}}%

\global\long\def\DissSpc{{\rm Diss}_{\mathfrak{sp}_{c}}}%
\global\long\def\SDissSpc{{\rm SDiss}_{\mathfrak{sp}_{c}}}%

\global\long\def\llangle{\langle\!\langle}%
{} 
\global\long\def\rrangle{\rangle\!\rangle}%
{} 
\global\long\def\rara{\rightrightarrows}%

\global\long\def\image{\textup{im}\,}%
 
\global\long\def\Be{\mathsf{Be}}%
\global\long\def\Lagr{\textup{Lagr}}%
\global\long\def\bSp{{\bf Sp}}%
\global\long\def\We{\mathsf{We}}%
 
\global\long\def\be{\mathsf{be}}%

\global\long\def\bF{{\bf F}}%
\global\long\def\sfb{\mathsf{b}}%
\global\long\def\sfB{\mathsf{B}}%
\global\long\def\sfS{\mathsf{S}}%
 
\global\long\def\ora#1{\overrightarrow{#1}}%

{} 
\global\long\def\graph{\textup{graph}}%
 
\global\long\def\eps{{\bf r}}%

\renewcommand{\labelenumi}{(\arabic{enumi})}
\begin{abstract}
The main aim of this article is to show some intimate relations among
the following three notions: (1) the metaplectic representation of
$Sp(2n,\mathbb{R})$ and its extension to some semigroups, called
the Olshanski semigroup for $Sp(2n,\mathbb{R})$ or Howe's oscillator
semigroup, (2) antinormally-ordered quantizations on the phase space
$\mathbb{R}^{2m}\cong\mathbb{C}^{m}$, (3) path integral quantizations
where the paths are on the phase space $\mathbb{R}^{2m}\cong\mathbb{C}^{m}$.
In the Main Theorem, the metaplectic representation $\rho(e^{X})$
($X\in\mathfrak{sp}(2n,\R)$) is expressed in terms of generalized
Feynman--Kac(--It\^{o}) formulas, but in real-time (not imaginary-time)
path integral form. Olshanski semigroups play the leading role in
the proof of it.
\end{abstract}

\section{Introduction}

{} The main aim of this article is to show some intimate relations among
the following three notions:
\begin{enumerate}
\item the metaplectic representation of $\Sp(2n,\R)$ and its extension
to some semigroups, called e.g.\ the Olshanski semigroup for $\Sp(2n,\R)$
\cite{Nee00}, the symplectic category \cite{Ner11} or Howe's oscillator
semigroup~\cite{Fol89}.
\item antinormally-ordered quantizations on the phase space $\R^{2m}\cong\C^{m}$.%
\item path integral quantizations where the paths are on the phase space
$\R^{2m}\cong\C^{m}$.
\end{enumerate}
A very rough idea of the theory of general Olshanski semigroups is
as follows \cite{Nee00}. Let $G$ be a real Lie group, $\fg$ the
Lie algebra of $G$, and $G_{\C}$, $\fg_{\C}$ be their complexifications,
respectively. Let $\pi$ be a continuous unitary representation of
$G$ on a Hilbert space $\cH$, and $\d\pi$ be the corresponding
representation of $\fg_{\C}$ on $\cH^{\infty}$ (the space of smooth
vectors in $\cH$), where $\im\cdot\d\pi(X)$ is essentially self-adjoint
for all $X\in\fg$. Define the subsemigroup $\Gamma_{G,\pi}$ of $G_{\C}$
by
\[
\Gamma_{G,\pi}:=\left\{ ge^{X}|g\in G,X\in\im\fg,\d\pi(X)\text{ is bounded from above}\right\} ,
\]
which is called an Olshanski semigroup. Then the unitary representation
$\pi$ of $G$ is extended to the bounded representation $\hat{\pi}$
of $\Gamma_{G,\pi}$ by 
\[
\hat{\pi}(ge^{X})=\pi(g)e^{\d\pi(X)},\qquad g\in G,X\in\im\fg.
\]
If $\pi$ is a highest weight representations of $G$, then $\hat{\pi}$
turns out to be a holomorphic representation of $\Gamma_{G,\pi}$.
Furthermore, $\hat{\pi}$ turns out to be ``natural'': $\hat{\pi}$
can be seen as the maximal analytic continuation of $\pi$ in a sense.
Thus it is expected that the problems on highest weight representations
of $G$ is translated to those on holomorphic representations of $\Gamma_{G,\pi}$.
However, since the above definition of $\Gamma_{G,\pi}$ depends on
the representation $\pi$, we need to know $\pi$ to know $\Gamma_{G,\pi}$.
Thus we need some redefinition of Olshanski semigroups so as not to
refer to any representation of $G$ to do that translation.%
{} In fact, such a redefinition is possible:%
{} The definition of an Olshanski semigroup $\Gamma_{G}(W)$ in \cite{Nee00}
does not depend on $\pi$, but on an invariant cone $W$ in $\fg$.

Although our Main Theorem has no reference to the Olshanski semigroups,
they play the leading role in the proof of it.

The Main Theorem is outlined as follows. Let $m\in\N$. %
{} %
Define $\alpha_{k}:\R^{2m}\to\R\ (k=1,...,2m)$ by
\[
\alpha_{k}(\vec{x}):=x_{m+k},\qquad\alpha_{m+k}(\vec{x}):=-x_{k},\qquad k=1,...,m,\ \vec{x}:=(x_{1},...,x_{2m}),
\]
and the (positive-definite) \termi{magnetic Laplacian} $\Delta^{\alpha}$
on $\R^{2m}$ by 

\[
\Delta^{\alpha}:=-\sum_{k=1}^{2m}\left(\frac{\di}{\di x_{k}}+\im\alpha_{k}\right)^{2}.
\]
The symplectic group $\Sp(2m,\R)$%
{} acts on the phase space (symplectic vector space) $\R^{2m}\cong\C^{m}$
as linear symplectic transformations. Hence each $\cA\in\sp(2m,\R)$
(the Lie algebra of $\Sp(2m,\R)$)%
{} corresponds to a Hamiltonian function $H_{\cA}:\C^{m}\to\R$ which
generates a one-parameter group of linear symplectic transformations
on $\C^{m}$. It is explicitly given by
\begin{equation}
H_{\cA}(\vec{z}):=\frac{\im}{2}{\bf z}^{*}\cI\cA_{c}{\bf z},\qquad\cI:={\begin{pmatrix}I & O\\
O & -I
\end{pmatrix}},\quad\cA_{c}:=\cW\cA\cW^{-1},\quad\cW:=\frac{1}{\sqrt{2}}\begin{pmatrix}I & \im I\\
I & -\im I
\end{pmatrix},
\end{equation}
where ${\bf z}:=\left(\ol z_{1},....,\ol z_{m},z_{1},....,z_{m}\right)^{\TT},$
${\bf z}^{*}:=\left(z_{1},....,z_{m},\ol z_{1},....,\ol z_{m}\right).$

For any Hamiltonian function $H:\C^{m}\to\R$ which satisfies some
regularity conditions, let $\cE_{b}(H)$ denote the antinormally ordered
quantization of $H$, which is a self-adjoint operator on a Hilbert
subspace $\cH$ of $L^{2}(\C^{m})$ (see Sec.~\ref{sec:Antinormally-ordered-quantizatio}
and Sec.~\ref{sec:diffop}).

Then we have the following relation between the classical Hamiltonian
$H_{\cA}$ and the quantum Hamiltonian $\cE_{b}(H_{\cA})$:

\begin{shaded}%
\begin{thm}
(Cor.~\ref{thm:slimexphDelta}) For any $\cA\in\sp(2m,\R)$%
{} and $t\in\R$, 
\begin{align}
\slim_{\nnn\to\infty}e^{\nnn m}\exp\left(-\im tH_{\cA}-\frac{\nnn}{2}\Delta^{\alpha}\right) & =e^{-\im t\cE_{b}(H_{\cA})}\qquad\text{on }\cH,\label{eq:slimexp0}
\end{align}
where $H_{\cA}$ in the l.h.s.~is viewed as a multiplication operator
on $L^{2}(\C^{m})$.
\end{thm}

\end{shaded}

Possibly this theorem itself is not a new result, %
because the l.h.s.~%
can be explicitly realized as a Gaussian integral operator, which
has been extensively studied in various fields; see e.g.~\cite{Fol89,Ner11}
and references therein. %
However, (hopefully) it has some new aspects: 
\begin{enumerate}
\item Our proof does not refer to any specific Gaussian integral operator
realizations; Instead we use more algebraic method of Olshanski semigroups
for symplectic groups. In fact, the main part of the proof in Sec.~\ref{sec:Results-for}
is on the \emph{finite-dimensional} matrices. Thus it is expected
that this method can be extended to ``non-Gaussian'' cases, in terms
of more general Lie groups and their highest weight representations.
\item Notice that the form of the l.h.s.\ of (\ref{eq:slimexp0}) (without
limit, i.e.~with each fixed $\nu>0$) is suitable to the application
of the Feynman--Kac--It\^o formula (e.g.\ \cite{Sim79,LHB11,Gun10}).
Thus (\ref{eq:slimexp0}) leads to a ``real-time'' (not ``imaginary-time'')
path-integral representation of the ``quantum time evolution'' in
the r.h.s.
\item Also note that we can interpret the magnetic Laplacian geometrically:
$\Delta^{\alpha}=\nabla^{*}\nabla$, where $\nabla$ is a connection
on a complex line bundle over the phase space $\C^{m}$. For further
geometric implication, see \cite{Yam22a}.
\end{enumerate}
There has been extensive literature in rigorous justification of the
path-integral methods in quantum physics. However, it appears that
there are only few rigorous studies on \emph{real-time geometric}
path integrals. (Note that the general Feynman--Kac formula for vector
bundles on Riemannian manifolds is formulated and proven relatively
recently; see G\"{u}neysu \cite{Gun10}.) One of our%
{} ``real-time geometric path integral formulas'' is stated as follows:

\begin{shaded}%
\begin{thm}
(a rough outline of Thm.~\ref{thm:vacextPathInt-bdd} and Cor.~\ref{thm:vacextPathInt})
There exists a sequence of probability measures $\mu_{\nnn},\nnn\in\N$,
(in fact, Brownian bridge measures) on the loop space
\[
{\rm loop}_{0}:=\left\{ \varphi\in C([0,1],\R^{2m})|\varphi(0)=\varphi(1)=0\right\} 
\]
such that for any $\cA\in\sp(2m,\R)$%
, the ``vacuum expectation value'' of $e^{-\im\cE_{b}(H_{\cA})}$
is expressed as follows:
\begin{align}
\bigl\langle\Omega_{0}|e^{-\im\cE_{b}(H_{\cA})}\Omega_{0}\bigr\rangle=\lim_{\nnn\to\infty}e^{\nnn m}\int_{{\rm loop}_{0}}e^{\im S_{\cA}(\varphi)}\d\mu_{\nnn}(\varphi)=\lim_{\nnn\to\infty}e^{\nnn m}\Ex_{\nnn}\Bigl[e^{\im S_{\cA}(\varphi)}\Bigr]\label{eq:vacexp=00003Dexp}
\end{align}
where
\[
S_{\cA}(\varphi):=\int_{\varphi}\alpha^{\flat}+\int_{0}^{1}H_{\cA}(\varphi(t))\d t,\qquad\alpha^{\flat}:=\sum_{k=1}^{2m}\alpha_{k}\d x_{k}.
\]
Here $\Ex_{\nnn}[\cdot]$ denotes the \emph{(classical)} expectation
value w.r.t.~the probability measure $\mu_{\nnn}$.
\end{thm}

\end{shaded}

In \cite{Yam22a}, we proved some similar results for the cases where
the Hamiltonian is bounded, and for more general phase spaces: (possibly
non-compact) complete K\"ahler manifolds satisfying some technical
conditions. Thus the above theorem is an extension of the result of
\cite{Yam22a}. 
\begin{rem}
Let $G$ be a locally compact group, and $\pi$ a unitary representation
of $G$ on $\cH$. For $v\in\cH\setminus\{0\}$, define $\phi_{v}:G\to\C$
by $\phi_{v}(g):=\langle v|\pi(g)v\rangle$. Then $(\pi,\cH)$ can
be reconstructed from $\phi_{v}$ by the GNS construction. If $G$
is a connected Lie group and $\phi_{v}$ is real analytic, then the
restriction $\phi_{v}\upha U$ of $\phi_{v}$ for some neighborhood
$U$ of the unit suffices to reconstruct $(\pi,\cH)$, in terms of
the GNS together with the analytic continuation.
\end{rem}

\begin{rem}
Eq.~(\ref{eq:vacexp=00003Dexp}) may have some conceptual implications
because it says that a quantum expectation value can be approximated
by some expectation values in the classical sense, not the converse.
It also suggests the Monte Carlo methods for numerical computations
of quantum expectation values, while they do not seem very effective.
\end{rem}

\begin{rem}
In the above theorem, the ``quantum Hamiltonian'' $\cE_{b}(H_{\cA})$
is a generator of a projective unitary representation of $\Sp(2m,\R)$,
rather than a quantum Hamiltonian of some realistic physical system.
Hence the physical meaning of this path-integral representation is
not so clear, but mathematically this may be intriguing since $\cE_{b}(H_{\cA})$
is non-semibounded, i.e.~unbounded both from below and from above,
in general; in such cases, most of the conventional Feynman--Kac
methods of imaginary-time path integrals will not be applicable since
both $e^{-t\cE_{b}(H_{\cA})}$ and $e^{t\cE_{b}(H_{\cA})}$ ($t>0$)
are unbounded.
\end{rem}

\section{Symplectic groups and Olshanski semigroups}

In this section we give basic definitions on symplectic groups and
related topics \cite{Ner11,Fol89}; For the general theory of Olshanski
semigroups, see \cite{Nee00}. The symbols defined in this section%
{} are mainly taken from Neretin \cite{Ner11}, and partially from Folland
\cite{Fol89}. However note that the term ``Olshanski semigroup''
is not found in \cite{Ner11,Fol89}, but in Neeb \cite{Nee00}, while
precisely he spells ``Ol'shanski\u{\i}''; Neretin \cite{Ner11}
use the term ``contraction semigroup'', which seems slightly vague
and misleading.

Define $\cJ\in\Mat(2n,\R)$ by 
\[
\SYM{\cJ}J\equiv\SYM{\cJ_{2n}}{J2n}:=\begin{pmatrix}O_{n} & I_{n}\\
-I_{n} & O_{n}
\end{pmatrix},
\]
where $O_{n}$ and $I_{n}$ are the $n\times n$ zero matrix and the
identity matrix, respectively. The symplectic form on $\R^{2n}$ %
{} determined by $\cJ$ is written by $\omega$: $\omega(u,v):=u^{\TT}\cJ v$
for $u,v\in\R^{2n}$ (as column vectors). Let $\bbK=\R,\C$, and consider
the symplectic group over $\bbK$: 
\begin{align*}
\SYM{\Spn{\bbK}}{Sp(n,K)}: & =\left\{ \cA\in\GL(2n,\bbK)|\cA^{\TT}\cJ\cA=\cJ\right\} \\
 & =\left\{ \begin{pmatrix}A & B\\
C & D
\end{pmatrix}\in\GL(2n,\bbK)\Big|AB^{\TT}=BA^{\TT},\ CD^{\TT}=DC^{\TT},\ AD^{\TT}-BC^{\TT}=I_{n}\right\} .
\end{align*}
together with its Lie algebra 
\begin{align*}
\spn{\bbK} & =\left\{ \cA\in\Mat(2n,\bbK)|\cJ A+\cA^{\TT}\cJ=0\right\} \\
 & =\left\{ \begin{pmatrix}A & B\\
C & -A^{\TT}
\end{pmatrix}\Big|A,B,C\in\Mat(n,\bbK),\ B^{\TT}=B,\ C^{\TT}=C\right\} .
\end{align*}

Let
\[
\SYM{\cW}W\equiv\SYM{\cW_{2n}}{W2n}:=\frac{1}{\sqrt{2}}\begin{pmatrix}I_{n} & \im I_{n}\\
I_{n} & -\im I_{n}
\end{pmatrix},
\]
and for any $\cA\in\Mat(2n,\C)$, let 
\[
\SYM{\cA_{c}}{Ac}:=\cW\cA\cW^{-1}.
\]
Then we see $\cJ_{c}:=\cW\cJ\cW^{-1}={\scriptsize\begin{pmatrix}-\im I_{n} & 0\\
0 & \im I
\end{pmatrix}}.$ Let
\[
\SYM{\cI}I\equiv\SYM{\cI_{2n}}{I2n}:=\begin{pmatrix}-I_{n} & 0\\
0 & I_{n}
\end{pmatrix}=-\im\cJ_{c},
\]
and define the Hermitian form (indefinite inner product) 
\[
\left\langle u|v\right\rangle _{\cI}:=\left\langle u|\cI v\right\rangle =-\im u^{*}\cJ_{c}v,\qquad u,v\in\C^{2n}.
\]
Then the group $\rU(n,n)$ is defined by 
\begin{align*}
\SYM{\rU(n,n)}{U(n,n)} & =\left\{ g\in\Mat(2n,\C):\left\langle gu|gv\right\rangle _{\cI}=\left\langle u|v\right\rangle _{\cI}\text{ for all }u,v\in\C^{2n}\right\} \\
 & =\left\{ g\in\Mat(2n,\C):g^{*}\cI g=\cI\right\} 
\end{align*}
The group $\Spcn{\R}\,(\cong\Spn{\R})$ has several equivalent definitions:

\begin{align*}
\SYM{\Spcn{\R}}{Spc(n,R)}: & =\left\{ \cW\cA\cW^{-1}|\cA\in\Spn{\R}\right\} \\
 & =\left\{ \begin{pmatrix}P & Q\\
\ol Q & \ol P
\end{pmatrix}\Big|P,Q\in\Mat(n,\C),\ PP^{*}-QQ^{*}=I_{n},\ PQ^{\TT}=QP^{\TT}\right\} \\
 & =\rU(n,n)\cap\left\{ \begin{pmatrix}P & Q\\
\ol Q & \ol P
\end{pmatrix}\Big|P,Q\in\Mat(n,\C)\right\} \\
 & =\rU(n,n)\cap\Spn{\C}
\end{align*}
Thus any element of $\Spcn{\R}$ preserves both the bilinear form
$(v,w)\mapsto v^{\TT}\cJ w$ and the hermitian form $(v,w)\mapsto v^{*}\cI w$
on $\C^{2n}\times\C^{2n}$. Its Lie algebra is given by
\begin{equation}
\SYM{\spcn{\R}}{spc}=\fraku(n,n)\cap\spcn{\C}=\left\{ \begin{pmatrix}A & B\\
\ol B & \ol A
\end{pmatrix}\Big|A,B\in\Mat(n,\C),\ A^{*}=-A,\ B^{\TT}=B\right\} .\label{eq:def:spc}
\end{equation}
Let
\begin{align*}
\SYM{\Gamma\rU(n,n)}{GammaU}: & =\left\{ g\in\GL(2n,\C):\ \forall v\in\C^{2n},\ \left\langle gv|gv\right\rangle _{\cI}\le\left\langle v|v\right\rangle _{\cI}\right\} \\
 & =\left\{ g\in\GL(2n,\C):\ \cI-g^{*}\cI g\ge0\right\} ,\\
\SYM{\Gamma{\rm Sp}_{c}(2n)}{GammaSpJ}: & ={\rm Sp}(2n,\C)\cap\Gamma\rU(n,n).
\end{align*}
Evidently $\Gamma\rU(n,n)$ (resp.\ $\Gamma{\rm Sp}_{c}(2n)$) is
a semigroup. This is called a \termi{contraction semigroup} or \termi{Olshanski semigroup}
for $\rU(n,n)$ (resp.\ $\Spcn{\R}$).

Note that the Lie algebra of $\rU(n,n)$ is given by
\[
\SYM{\fraku(n,n)}{u(n,n)}:=\left\{ X\in\Mat(2n,\C)|\ \langle v|Xv\rangle_{\cI}\in\im\R\text{ for all }v\right\} .
\]
Let
\[
\SYM{\Diss(n,n)}{Diss()}:=\left\{ X\in\Mat(2n,\C)|\ \Re\langle v|Xv\rangle_{\cI}\leq0\text{ for all }v\right\} .
\]
\begin{align*}
\SYM{\SDiss(n,n)}{SDiss}: & =\left\{ X\in\Mat(2n,\C)|\ \langle v|Xv\rangle_{\cI}\leq0\text{ for all }v\right\} .\\
 & =\left\{ X\in\im\cdot\fraku(n,n)|\ \langle v|Xv\rangle_{\cI}\leq0\text{ for all }v\right\} \\
 & =\left\{ X\in\im\cdot\fraku(n,n)|\cI X\leq0\right\} \\
 & =\im\cdot\fraku(n,n)\cap\Diss(n,n)
\end{align*}
$\Diss(n,n)$ is called the cone of \termi{$\cI$-dissipative} operators.
$\SDiss(n,n)$ is called the cone of \termi{$\cI$-self-adjoint dissipative}
operators. Let%
\begin{align*}
\SYM{\DissSpc(2n)}{dissspc}: & =\spcn{\C}\cap\SDiss(n,n).\\
 & =\left\{ X\in\spcn{\C}|\Re\langle v|Xv\rangle_{\cI}\leq0\text{ for all }v\right\} 
\end{align*}
\begin{align*}
\SYM{\SDissSpc(2n)}{sdisssp}: & =\im\cdot\spcn{\R}\cap\SDiss(n,n)\\
 & =\im\cdot\spcn{\R}\cap\Diss(n,n)\\
 & =\left\{ X\in\im\cdot\spcn{\R}|\forall v,\ \Re\langle v|Xv\rangle_{\cI}\leq0\right\} \\
 & =\left\{ X\in\im\cdot\spcn{\R}|\cI X\leq0\right\} .
\end{align*}
We have%
\[
\Diss(n,n)=\fraku(n,n)\oplus\SDiss(n,n).
\]
\[
\DissSpc(2n)=\sp(2n,\R)\oplus\SDissSpc(2n).
\]

\begin{shaded}%
\begin{prop}
{\rm{}\cite[Theorem 7.5 in Ch.2.]{Ner11}} The following conditions
are equivalent:

\begin{enumerate}
\item $e^{tX}\in\Gamma\mathrm{U}(n,n)$ for all $t>0,$
\item $X\in\Diss(n,n)$. 
\end{enumerate}
\end{prop}

\end{shaded}

\begin{shaded}%
\begin{prop}
[The Potapov-Olshanski decomposition]\label{thm:3.5.2}%
{\rm{}(\cite{Ner11}: Theorem 7.7 in Ch.2 and Theorem 5.2 in Ch.3)}
Each element $g\in\Gamma\Sp_{c}(2n)$ admits a unique decomposition%
\[
g=he^{X},\qquad h\in\Spcn{\R},\ X\in\SDissSpc(2n).
\]
\end{prop}

\end{shaded}

\section{Metaplectic representation}

Let $a_{k},a_{k}^{*}$ ($k=1,...,n$) be annihilation/creation operators
on $\cH$ s.t.

\begin{equation}
[a_{k},a_{l}]=0,\qquad[a_{k},a_{l}^{*}]=\delta_{kl}.\label{eq:CCR}
\end{equation}
For $\cA\in\spcn{\R}$, define the (essentially) skew-self-adjoint
operator $\d\rho(\cA)$ by%
\begin{equation}
\d\rho(\cA):=\frac{1}{2}{\bf a}^{*}\cI\cA{\bf a}\label{eq:def:drhoa}
\end{equation}
\[
{\bf a}:=\left(a_{1}^{*},....,a_{n}^{*},a_{1},....,a_{n}\right)^{\TT}=\begin{pmatrix}\vec{a^{*}}\\
\vec{a}
\end{pmatrix},
\]
\[
{\bf a}^{*}:=\left(a_{1},....,a_{n},a_{1}^{*},....,a_{n}^{*}\right)=\left(\vec{a}^{\TT},\vec{a^{*}}^{\TT}\right).
\]
For example, we have
\[
\sp_{c}(2,\R)=\left\{ \begin{pmatrix}\im r & z\\
\ol z & -\im r
\end{pmatrix}\Big|r\in\R,\ z\in\C\right\} 
\]
and
\[
\d\rho\left(\begin{pmatrix}\im r & z\\
\ol z & -\im r
\end{pmatrix}\right)=\frac{\im}{2}\left(r\left(a_{1}a_{1}^{*}+a_{1}^{*}a_{1}\right)-\im\left(za_{1}^{2}-\ol za_{1}^{*2}\right)\right).
\]
Especially,
\[
\d\rho\left(\begin{pmatrix}\im & 0\\
0 & -\im
\end{pmatrix}\right)=\frac{\im}{2}\left(a_{1}a_{1}^{*}+a_{1}^{*}a_{1}\right)=\im\left(a_{1}^{*}a_{1}+\frac{1}{2}\right).
\]

\begin{lyxgreyedout}
\hidec{

\begin{shaded}%
\[
\sp_{c}(2,\R)=\left\{ \begin{pmatrix}A & B\\
\ol B & \ol A
\end{pmatrix}|A\in\im\R,\ B\in\C\right\} =\left\{ \begin{pmatrix}A & B\\
\ol B & -A
\end{pmatrix}|A\in\im\R,\ B\in\C\right\} =\left\{ \begin{pmatrix}ir & z\\
\ol z & -ir
\end{pmatrix}|r\in\R,\ z\in\C\right\} 
\]
\[
\im\sp_{c}(2,\R)=\left\{ \begin{pmatrix}A & iB\\
i\ol B & -A
\end{pmatrix}|A\in\R,\ B\in\C\right\} 
\]

\[
\SDissSpc(2)=\im\cdot\sp_{c}(2,\R)\cap\SDiss(1,1)=\left\{ \begin{pmatrix}r & iz\\
i\ol z & -r
\end{pmatrix}|\ r\in\R,\ z\in\C,\ r\le|z|\right\} 
\]
\end{shaded}
\begin{proof}
\end{proof}
\rule[0.5ex]{1\columnwidth}{1pt}

\begin{shaded}%
\[
\d\rho\begin{pmatrix}i & 0\\
0 & -i
\end{pmatrix}=-\frac{i}{2}\left(aa^{*}+a^{*}a\right)
\]
\end{shaded}
\begin{proof}
\end{proof}
\rule[0.5ex]{1\columnwidth}{1pt}

\begin{shaded}%
\[
\d\rho\begin{pmatrix}ir & z\\
\ol z & -ir
\end{pmatrix}=-\frac{1}{2}\left(ir\left(aa^{*}+a^{*}a\right)+zaa-\ol za^{*}a^{*}\right)
\]
\end{shaded}
\begin{proof}
\end{proof}
\rule[0.5ex]{1\columnwidth}{1pt}

}%
\end{lyxgreyedout}

We find that $\d\rho$ is faithful representation of $\spcn{\R}$
on the space $\cH^{\infty}$ of smooth vectors of $\cH$. Roughly
speaking, the {metaplectic representation} of $\Spcn{\R}$ is determined
by this representation of $\spcn{\R}$; The set of unitary operators
$\left\{ e^{\d\rho(\cA)}|\cA\in\spcn{\R}\right\} $ generates a group
$\SYM{\Mpn{\R}}{Mp(n,R)}\subset\rU(\cH)$ which satisfies $\Mpn{\R}/\{\pm I\}\cong\Spn{\R}$.
Hence it determines a projective unitary representation $\rho$ of
$\Spcn{\R}\cong\Spn{\R}$, called the \termi{metaplectic representation}
(also called the oscillator, harmonic, or Segal--Shale--Weil representation). 

The double-valuedness of the metaplectic representation is not very
important in this article; Here we take account only of the projectivity
of it. Hence we will consider as follows. Let $\SYM{\Bdd(\cH)}{B(H)}$
denote the semigroup of bounded operators on $\cH$. Then $\Bdd(\cH)/\C^{\times}$
($\C^{\times}:=\C\setminus\{0\}$) also is a semigroup. More precisely
we consider the semigroup $\left(\Bdd(\cH)\setminus\{0\}\right)/\C^{\times}$,
which has no zero element. For each $A\in\Bdd(\cH)\setminus\{0\}$,
$\tilde{A}:=\C^{\times}A\in\Bdd(\cH)/\C^{\times}$ is called an \termi{operator ray}
(of $A$).

The norm (resp. strong, weak) topology on $\Bdd(\cH)/\C^{\times}$
is naturally determined by the norm (resp.\ strong, weak) topology
on $\Bdd(\cH)$. Let $A_{k}\in\Bdd(\cH)\setminus\{0\}$ ($k=1,2,...,\infty$),
and $\tilde{A}_{k}:=\C^{\times}A_{k}\in\Bdd(\cH)/\C^{\times}$. Then,
$\tilde{A}_{k}\to\tilde{A}_{\infty}$ as $k\to\infty$ in norm (resp.\ strong,
weak) topology iff there exists a sequence $z_{1},z_{2},...\in\C^{\times}$
s.t. $z_{k}A_{k}\to A_{\infty}$ in norm (resp.\ strong, weak) topology.

Let $G$ be a topological semigroup. A strongly continuous homomorphism
$\pi:G\to\Bdd(\cH)/\C^{\times}$ is called a (continuous) \termi{projective representation}
of $G$. If $U\in\Bdd(\cH)$ is unitary, we also call $\C^{\times}U\in\Bdd(\cH)/\C^{\times}$
\termi{unitary}. The set of unitary element of $\Bdd(\cH)/\C^{\times}$
is denoted by $\rU(\cH)/\C^{\times}$, while a more precise notation
is $\left(\C^{\times}\rU(\cH)\right)/\C^{\times}$ . If $G$ is a
group and $\pi:G\to\rU(\cH)/\C^{\times}$ is a homomorphism, we call
$\pi$ a \termi{projective unitary representation} (PUR). We regard
the metaplectic representation $\rho$ as the projective unitary representation
$\rho:\Spcn{\R}\to\rU(\cH)/\C^{\times}$ s.t. $\rho(e^{X})=\widetilde{e^{\d\rho(X)}}$,
$X\in\spcn{\R}$. 

The definition (\ref{eq:def:drhoa}) of $\d\rho(X)$ is naturally
generalized. First, if $X\in\SDissSpc(2n)$, we find that $\d\rho(X)$
is defined as an essentially self-adjoint operator bounded from above.
Hence $e^{\d\rho(X)}$ is also defined as a bounded operator. 

By Theorem \ref{thm:3.5.2}, the metaplectic representation $\rho$
is extended to a projective representation $\tilde{\rho}$ of the
Olshanski semigroup $\Gamma\Sp_{c}(2n)$ as follows. Let $g\in\Gamma\Sp_{c}(2n)$,
and $g=he^{X}$, $h\in\Spcn{\R},\ X\in\SDissSpc(2n).$ Then we define
$\hat{\rho}(g)$ by
\[
\hat{\rho}(g):=\rho(h)e^{\d\rho(X)}.
\]
In the following, we write $\rho(g)$ for $\hat{\rho}(g)$.

\begin{lyxgreyedout}
\hidec{

Consider the subsemigroup $\SYM{\osgn}O$ of $\Bdd(\cH)/\C^{\times}$
{} generated by the set of bounded operators%

\[
\left\{ \C^{\times}e^{\alpha\d\rho(\cA)}|\cA\in\spcn{\R},\ \d\rho(\cA)\text{ is bounded from above},\ \alpha\in\C,\ \Re\alpha\ge0\right\} .
\]
{} Moreover we find%
\[
\osgn=\left\{ \C^{\times}Ue^{\im\cdot\d\rho(\cA)}|U\in\Mpn{\R},\ \cA\in\spcn{\R},\ \im\cdot\d\rho(\cA)\text{ is bounded from above}\right\} .
\]
We call $\osgn$ the \termi{oscillator semigroup}. %
{} Correspondingly, define the subsemigroup $\GSpcn$ of $\Spcn{\C}$,
called the \termi{Olshanski semigroup} (w.r.t.\ the metaplectic
representation $\rho$), by
\[
\SYM{\GSpcn}{GammaSp}:=\left\{ ge^{\im\cA}|g\in\Spcn{\R},\ \cA\in\spcn{\R},\ \im\cdot\d\rho(\cA)\text{ is bounded from above}\right\} .
\]
It is shown that the metaplectic representation $\rho$ of $\Spn{\R}$
has a holomorphic extension to the a projective representation of
$\GSpn$, so that
\[
\rho\left(ge^{\im\cA}\right)=\rho(g)e^{\im\cdot\d\rho(\cA)},\qquad g\in\Spcn{\R},\ \cA\in\spcn{\R}.
\]
These properties follows from general theorems for highest weight
representations (see Theorem X.3.9 and Theorem XI.2.3 in \cite{Nee00}).
See also \cite[Ch.5]{Fol89} and \cite[Sec.2.7, Sec.3.5]{Ner11}. 

\rule[0.5ex]{1\columnwidth}{1pt}

\begin{lyxgreyedout}
\hidec{

The set of unitary operators

\[
\left\{ \exp\left(\im P_{\cA}(\vec{a},\vec{a}^{*})\right)|\cA\in\spcn{\R}\right\} 
\]
generates a group $\SYM{\Mp(n,\R)}{Mp(n,R)}\subset\rU(\cH)$ which
satisfies $\Mp(n,\R)/\{\pm I\}\cong\Spn{\R}$. Hence it determines
a projective unitary representation $\rho$ of $\Spn{\R}$, called
the \termi{metaplectic representation} (also called the oscillator,
harmonic, or Segal--Shale--Weil representation).

Consider the subsemigroup $\SYM{\cO_{n}}O$ of $\Bdd(\cH)/\C^{\times}$
generated by the set of bounded operators
\[
\left\{ \C^{\times}\exp\left(\alpha P_{\cA}(\vec{a},\vec{a}^{*})\right)|\cA\in\spcn{\R},\ P_{\cA}(\vec{a},\vec{a}^{*})\text{ is bounded from above},\ \alpha\in\C,\ \Re\alpha\ge0\right\} .
\]
{} Moreover we find
\[
\cO_{n}=\left\{ \C^{\times}Ue^{P_{\cA}(\vec{a},\vec{a}^{*})}|U\in{\rm Mp}(n,\R),\ \cA\in\spcn{\R},\ P_{\cA}(\vec{a},\vec{a}^{*})\text{ is bounded from above}\right\} .
\]
This follows from general theorems for highest weight representations
(see Theorem X.3.9 and Theorem XI.2.3 in \cite{Nee00}; see also \cite[Ch.5]{Fol89}
and \cite[Sec.2.7, Sec.3.5]{Ner11}). We call $\cO_{n}$ the \termi{oscillator semigroup}.
{} Correspondingly, define the subsemigroup $\GSpn$ of $\Spn{\C}$,
called the \termi{Olshanski semigroup} (w.r.t.\ the metaplectic
representation $\rho$), by
\[
\SYM{\GSpcn}{GammaSpc}:=\left\{ ge^{-\im\cA}|g\in\Spcn{\R},\ \cA\in\spcn{\R},\ P_{\cA}(\vec{a},\vec{a}^{*})\text{ is bounded from above}\right\} .
\]
It is shown that the metaplectic representation $\rho$ of $\Spcn{\R}$
has a holomorphic extension to the a projective representation of
$\GSpcn$, so that
\[
\rho\left(ge^{-\im\cA}\right)=\rho(g)e^{P_{\cA}(\vec{a},\vec{a}^{*})},\qquad g\in\Spcn{\R},\ \cA\in\spcn{\R}.
\]

}%
\end{lyxgreyedout}

\begin{rem}
We defined $P_{\cA}(\vec{a},\vec{a}^{*})$ by the Weyl quantization.
However, even if instead we take the nomally-ordered or antinormally-ordered
quantization, we get the same definitions of $\cO_{n}$ and $\GSpcn$.
\end{rem}

}%
\end{lyxgreyedout}

\begin{rem}
Roughly speaking, Howe's \termi{oscillator semigroup} \cite{Fol89}
is given by $\rho(\Gamma\Sp_{c}(2n))$, the range of the metaplectic
representation of $\Gamma\Sp_{c}(2n)$. However, it is not precise;
According to the definition in \cite{Fol89}, any element of the oscillator
semigroup is a Hilbert--Schmidt operator. If $n\ge2$, the operator
$e^{-ta_{1}^{*}a_{1}}$ ($t\ge0$) is not Hilbert--Schmidt, and hence
not in the oscillator semigroup in the sense of \cite{Fol89}. On
the other hand, we see $\widetilde{e^{-ta_{1}^{*}a_{1}}}\in\rho(\Gamma\Sp_{c}(2n))$.
\end{rem}

\section{Antinormally-ordered quantization}

\label{sec:Antinormally-ordered-quantizatio}

{} Assume that $n$ is even, $n=2m$, %
{} and
\[
b_{k}:=a_{m+k},\qquad k=1,...,m
\]
\[
Z_{k}:=a_{k}^{*}+b_{k}\qquad k=1,...,m.
\]
Then we see 
\[
[Z_{k},Z_{l}]=[Z_{k},Z_{l}^{*}]=0,\qquad k,l=1,...,m,
\]
i.e.\ $\{Z_{k}\}_{k=1,...,m}$ are (unbounded) commuting normal operators.
Thus, for any Borel function $f:\C^{m}\to\C$, the operator $f(Z_{1},...,Z_{m})$
is well-defined.

Let

\[
\SYM{N_{a}}{Na}:=\sum_{k=1}^{m}a_{k}^{*}a_{k},\qquad\SYM{N_{b}}{Nb}:=\sum_{k=1}^{m}b_{k}^{*}b_{k},
\]
and $\SYM{E_{a}}{Ea}$ (resp.\ $\SYM{E_{b}}{Eb}$) be the orthogonal
projection onto the subspace $\ker N_{a}\subset\cH$ (resp.\ $\ker N_{b}\subset\cH$).
Let
\[
\SYM{\cE_{a}(X)}{Ea()}:=E_{a}XE_{a},\qquad\SYM{\cE_{b}(X)}{Eb()}:=E_{b}XE_{b}.
\]
Then we see
\begin{align*}
 & \cE_{a}(Z_{k}^{p}Z_{l}^{*q})=\cE_{a}(b_{k}^{p}b_{l}^{*q})=b_{k}^{p}b_{l}^{*q}E_{a},\\
 & \cE_{b}(Z_{k}^{p}Z_{l}^{*q})=\cE_{b}(a_{k}^{p}a_{l}^{*q})=a_{k}^{p}a_{l}^{*q}E_{b},\qquad k,l=1,...,m,\ p,q=0,1,...
\end{align*}
This means that the map 
\[
f\mapsto\cE_{a}(f(\vec{Z})),\quad\text{resp. }f\mapsto\cE_{b}(f(\vec{Z})),\qquad\vec{Z}:=(Z_{1},...,Z_{m})
\]
is the antinormally-ordered quantization of the function $f$ on the
phase space $\C^{m}\cong\R^{n}$ in terms of $\{b_{k}\}_{k=1,...,m}$
(resp.\ $\{a_{k}\}_{k=1,...,m}$). %

We expect that this ``quantization by projection $E_{a/b}$'' viewpoint
makes the notion of quantization more transparent. However, the nature
of the operation $\cE_{a}$ (or $\cE_{b}$) is not so clear, contrary
to the appearance. To examine the projections $E_{a},E_{b}$ further,
we express them by the creation/annihilation operators as follows

\[
E_{a}=\lim_{\nnn\to\infty}e^{-\nnn N_{a}},\qquad E_{b}=\lim_{\nnn\to\infty}e^{-\nnn N_{b}}.
\]

Furthermore, we wish to describe them in terms of the symplectic groups
and its metaplectic representations.

In the following, we consider $E_{b}$ and $N_{b}$ only. Define $\cN_{b}\in\Mat(2n,\C)$
by

\[
\SYM{\cN_{b}}{Nb}:=O_{m}\oplus I_{m}\oplus O_{m}\oplus(-I_{m})={\rm diag}(\overbrace{0,...,0}^{m},\overbrace{1,...,1}^{m},\overbrace{0,...,0}^{m},\overbrace{-1,...,-1}^{m}).
\]
We see $-\cN_{b}\in\SDissSpc(2n)\subset\im\cdot\spcn{\R}$, and%
\[
\d\rho\left(-\cN_{b}\right)=-\left(N_{b}+\frac{m}{2}\right).
\]
Hence%
\[
\tilde{E}_{b}=\lim_{\nnn\to\infty}\C^{\times}e^{\nnn\d\rho\left(-\cN_{b}\right)}=\lim_{\nnn\to\infty}\rho(e^{-\nnn\cN_{b}}).
\]
in norm (and also strong or weak) topology on $\Bdd(\cH)/\C^{\times}$.
(Note that for $A\in\Bdd(\cH)\setminus\{0\}$ and $z\in\C\setminus\{0\}$,
we find $\widetilde{A+zI}\neq\widetilde{A}$ but $\widetilde{e^{A+zI}}=\widetilde{e^{A}}$,
where $\widetilde{X}:=\C^{\times}X$.) 

Thus our quantization procedure is related with the limit $\lim_{\nnn\to\infty}\rho(g_{\nnn})$
where $\left\{ g_{\nu}\right\} _{\nu\in\N}$ is some sequence in the
Olshanski semigroup $\Gamma\Sp_{c}(2n)$.%
{} If $g_{\infty}:=\lim_{\nnn\to\infty}g_{\nnn}$ exists in $\Gamma\Sp_{c}(2n)$,
we have $\slim_{\nnn\to\infty}\rho(g_{\nnn})=\rho(g_{\infty})$, and
hence it suffices to deal with the finite-dimensional matrix $g_{\infty}$.
However, for %
$g_{\nnn}:=e^{-\nnn\cN_{b}}$, we see $g_{\nnn}$ diverges as $\nnn\to\infty$
in $\Mat(2n,\C)$. We shall see that such divergent cases are of our
main concern, in the context of quantizations. An elegant solution
of this problem is given by the graph viewpoint \cite{Ner11}, which
will be explained in Sec.\,\ref{sec:graph}.

\section{Number operator representation of quantized time evolution}

Let $f:\C^{m}\to\R$ (a ``classical Hamiltonian''). The quantum
time evolution w.r.t.\ the quantized Hamiltonian $\cE_{b}(f(\vec{Z}))$
is given by the unitary operators $U_{f,t}:=\exp[-it\cE_{b}(f(\vec{Z}))]$,
$t\in\R$. The following conjecture expresses $U_{f,t}$ without the
projection $\cE_{b}$ but with $N_{b}$, the number operator which
is more familiar and usable for calculations.%
{} %

\begin{shaded}%
\begin{conjecture}
[Number operator representation of quantized time evolution]\label{conj:sibori}If
the classical Hamiltonian $f$ is ``physically reasonable,'' then
$\cE_{b}(f(\vec{Z}))$ is essentially self-adjoint, and the following
holds:
\begin{equation}
\lim_{\nnn\to\infty}\exp[-\im tf(\vec{Z})-\nnn N_{b}]\varphi=U_{f,t}\varphi,\qquad t\in\R,\ \varphi\in\ran E_{b}.\label{eq:sibori}
\end{equation}
in other words,
\[
\slim_{\nnn\to\infty}\exp[-\im tf(\vec{Z})-\nnn N_{b}]=U_{f,t}E_{b},\qquad t\in\R.
\]
\end{conjecture}

\end{shaded}

However, it does not seem easy to formulate the ``physical reasonability''
rightly and to prove (\ref{eq:sibori}). In fact, even the essential
self-adjointness of $\cE_{b}(f(\vec{Z}))$ is not assured in general.
(If $f$ is a classical Hamiltonian in the narrow sense, i.e.\ if
$f$ can be interpreted as the energy of the system, $f$ is bounded
from below typically, and hence the essential self-adjointness is
widely assured. However, we want to consider more general Hamiltonian
functions $f$, which are generators of various continuous physical
transformations, not only time translations; In such cases $f$ is
often unbounded both from below and from above.)

We verify Conjecture \ref{conj:sibori} in the case where $f:\C^{m}\cong\R^{2m}\to\R$
is the Hamiltonian function of $\spcN m{\R}$.

For $\cA\in\spcN m{\R}$, the corresponding Hamiltonian function $h_{\cA}:\C^{m}\to\C$
is given by

\begin{equation}
\SYM{h_{\cA}(\vec{z})}{hA}:=\frac{1}{2}{\bf z}^{*}\cI\cA{\bf z},\qquad\vec{z}:=(z_{1},...,z_{m}),\label{eq:def:hA}
\end{equation}
where
\[
{\bf z}:=\left(\ol z_{1},....,\ol z_{m},z_{1},....,z_{m}\right)^{\TT},\quad{\bf z}^{*}:=\left(z_{1},....,z_{m},\ol z_{1},....,\ol z_{m}\right).
\]
Indeed, we see the value of $h_{\cA}(\vec{z})$ is pure imaginary.
This seems strange since a classical-mechanical Hamiltonian function
on a phase space is usually real-valued. However we prefer to retain
the similarity to (\ref{eq:def:drhoa}). If one prefers the consistency
with classical mechanics, instead take $h_{\cA}'(\vec{z}):=\im h_{\cA}(\vec{z})$.

The following lemma is shown by a straightforward calculation.

\begin{shaded}%
\begin{lem}
\label{thm:hA=00003Ddrho}Let $\cA=\begin{pmatrix}A & B\\
\ol B & \ol A
\end{pmatrix}\in\spcN m{\R}$, so that $A,B\in\Mat(m,\C)$, \textup{$A^{*}=-A,\ B^{\TT}=B$. Then}
\begin{equation}
h_{\cA}(\vec{Z})=\frac{1}{2}{\bf Z}^{*}\cI_{2m}\cA{\bf Z}=\frac{1}{2}{\bf a}^{*}\cI_{2n}\hat{\cA}{\bf a}=\d\rho(\hat{\cA}),\qquad\vec{Z}:=(Z_{1},...,Z_{m}),\label{eq:hA=00003Ddrho}
\end{equation}
where
\[
{\bf Z}:=\left(Z_{1}^{*},....,Z_{m}^{*},Z_{1},....,Z_{m}\right)^{\TT},\qquad{\bf Z}^{*}:=\left(Z_{1},....,Z_{m},Z_{1}^{*},....,Z_{m}^{*}\right),
\]
\[
\hat{\cA}:=\begin{pmatrix}-\ol A & -\ol B & -\ol B & -\ol A\\
B & A & A & B\\
-B & -A & -A & -B\\
\ol A & \ol B & \ol B & \ol A
\end{pmatrix}\in\spcn{\R}.
\]
\end{lem}

\end{shaded}

Substitute $-\im tf:=h_{\cA}$ in the conjecture (\ref{eq:sibori}),
then we have
\begin{equation}
\slim_{\nnn\to\infty}\exp[h_{\cA}(\vec{Z})-\nnn N_{b}]=\exp[\cE_{b}(h_{\cA}(\vec{Z}))]E_{b}.\label{eq:sibosi1}
\end{equation}
Notice that (\ref{eq:hA=00003Ddrho}) implies
\[
h_{\cA}(\vec{Z})-\nnn N_{b}=\d\rho\left(\hat{\cA}-\nnn\cN_{b}\right)+\frac{\nnn m}{2}.
\]
Furthermore, consider the convergence as operator rays instead of
operators, i.e.\ the topology of $\Bdd(\cH)/\C^{\times}$, instead
of $\Bdd(\cH)$, then (\ref{eq:sibosi1}) becomes%
\begin{equation}
\slim_{\nnn\to\infty}\rho\Bigl(\exp\bigl(\hat{\cA}-\nnn\cN_{b}\bigr)\Bigr)=\C^{\times}\exp\left[\cE_{b}(h_{\cA}(\vec{Z}))\right]E_{b},\label{eq:siborihA}
\end{equation}
where the term $\nnn m/2$ vanishes. The first goal of this article
is to prove this. Note that let $g_{\nnn}:=\exp\bigl(\hat{\cA}-\nnn\cN_{b}\bigr)$,
then we see $g_{\nnn}\in\GSpcn$. Here we again encounter the problem
of convergence of the sequence $(g_{\nnn})$ in the Olshanski semigroup
$\GSpcn$.

\section{Extended Olshanski semigroup}

\label{sec:graph}

To deal with the convergence problem in the Olshanski semigroup $\GSpcn$,
we introduce the notion of extended Olshanski semigroup. Since this
notion does not seem common in the community of mathematical physics,
we outline the theory here. This section is based on Neretin \cite{Ner11}:
Especially, see Sec.\,8 and 9 %
in Ch.\,1, Sec.\,9 in Ch.\,2, and Sec.\,1 in Ch.\,5 of \cite{Ner11}.
Again note that the term ``(extended) Olshanski semigroup'' is not
found in \cite{Ner11}, but in Neeb \cite{Nee00}; Our notion of extended
Olshanski semigroup for $\Spcn{\R}$ amounts to the semigroup of morphisms
in the ``symplectic category'' in \cite{Ner11}.

{} Let $T\in\Mat(2n,\C)$ and regard it as a linear operator $T:\C^{2n}\to\C^{2n}$.
The graph of $T$
\[
\graph(T):=\left\{ v\oplus Tv|v\in\C^{2n}\right\} \subset\C^{2n}\oplus\C^{2n}
\]
is a $2n$-dimensional linear subspace of $\C^{2n}\oplus\C^{2n}\cong\C^{4n}$.
Thus $T$ is identified with a single point in the Grassmann manifold
$G_{4n,2n}(\C)$. 

Observe that $\graph(e^{-\nnn\cN_{b}})$ converges in $G_{4n,2n}(\C)$
as $\nnn\to\infty$, although $\lim_{\nnn\to\infty}e^{-\nnn\cN_{b}}$
does not converge to any matrix. If we can reformulate the metaplectic
representation $\rho(g)$ as $\rho(\graph(g))$ ($g\in\GSpcn$), it
is expected that
\[
\lim_{\nnn\to\infty}\rho\left(e^{-\nnn\cN_{b}}\right)=\rho\left(\lim_{\nnn\to\infty}\graph(e^{-\nnn\cN_{b}})\right).
\]
holds. However, the rhs is not defined yet. It will be found in Theorem
\ref{thm:Sp2n-rep} below.

For any subspaces $A,B\subset\C^{2n}\oplus\C^{2n}$, define the product
$AB\subset\C^{2n}\oplus\C^{2n}$ by
\[
AB:=\left\{ x\oplus y\in\C^{2n}\oplus\C^{2n}|\exists w\in\C^{2n},\ x\oplus w\in P,\ w\oplus y\in Q\right\} ,
\]
and $\ker A,\indef A\subset\C^{2n}$ by
\[
\ker A:=\left\{ x\in\C^{2n}|x\oplus0\in A\right\} ,\qquad\indef A:=\left\{ y\in\C^{2n}|0\oplus y\in A\right\} 
\]

\begin{defn}
Define $\SYM{{\bf U}_{n,n}}{Sp2n}$ to be the set of $2n$-dimensional
subspaces $P$ of $\C^{2n}\oplus\C^{2n}$ %
{} such that \renewcommand{\labelenumi}{(\arabic{enumi})}

\begin{enumerate}
\item $P$ is $\cI$-contractive, i.e. %
{} for any $v\oplus w\in P$, $\left\langle v|v\right\rangle _{\cI}\ge\left\langle w|w\right\rangle _{\cI}.$
\item %
for any nonzero $v\in\indef P$, $\left\langle v|v\right\rangle _{\cI}<0$.
\item for any nonzero $v\in\ker P$, $\left\langle v|v\right\rangle _{\cI}>0$.
\end{enumerate}
\end{defn}

\begin{defn}
Let $P$ be a subspace of $\C^{2n}\oplus\C^{2n}$. $P$ is said to
be \termi{symplectic} if for any $v\oplus w,\,v'\oplus w'\in P$,
$v^{\TT}\cJ v^{\prime}=w^{\TT}\cJ w^{\prime}$. Let
\[
\SYM{\bSp_{2n}}{Sp2n}:=\left\{ P\in{\bf U}_{n,n}|P\text{ is symplectic}\right\} ,
\]

\end{defn}

Since $\bSp_{2n}\subset{\bf U}_{n,n}\subset G_{4n,2n}(\C)$, ${\bf U}_{n,n}$
and $\bSp_{2n}$ are given the topology induced by that of $G_{4n,2n}(\C)$.

\begin{shaded}%
\begin{prop}
Let ${\bf G}$ be ${\bf U}_{n,n}$ or $\bSp_{2n}$. For any $A,B\in{\bf G}$,
we have $AB\in{\bf G}$; that is, ${\bf G}$ is a semigroup. The product
in ${\bf G}$ is continuous.
\end{prop}

\end{shaded}
\begin{proof}
See Sec.9.1 in Ch.1 of \cite{Ner11}. However, the continuity of $(A,B)\mapsto AB$
does not appear to be stated explicitly in \cite{Ner11}. It follows
from Proposition \ref{thm:Ner2.9.4} below.
\end{proof}
\begin{shaded}%
\begin{prop}
{} (1) If $g\in\Gamma\rU(n,n)$, then $\graph(g)\in{\bf U}_{n,n}$.
Furthermore the semigroup $\Gamma\rU(n,n)$ is embedded continuously
and densely into the semigroup ${\bf U}_{nn}$ by $g\mapsto\graph(g)$.

(2) If $g\in\Gamma\Sp_{c}(2n)$, then $\graph(g)\in\bSp_{2n}$. Furthermore
the semigroup $\Gamma\Sp_{c}(2n)$ is embedded continuously and densely
into the semigroup $\bSp_{2n}$ by $g\mapsto\graph(g)$. 
\end{prop}

\end{shaded}
\begin{proof}
Recall that $g\in\Gamma\rU(n,n)$ iff $\left\langle gv|gv\right\rangle _{\cI}\le\left\langle v|v\right\rangle _{\cI}$,%
{} and $\Gamma{\rm Sp}_{c}(2n)={\rm Sp}(2n,\C)\cap\Gamma\rU(n,n)$.
{} Then the above is evident except the density, which is not stated
explicitly in \cite{Ner11}. The density follows from Proposition
\ref{thm:2.8.1} below.
\end{proof}
We call $\bSp_{2n}$ the \termi{extended Olshanski semigroup}, or
the \termi{Krein--Shmul'yan--Olshanski (KSO) semigroup} for $\Spcn{\R}$.

Let $V_{\pm}:=\ker(\cI\mp1)\subset\C^{2n}$, so that $\C^{2n}=V_{-}\oplus V_{+}$.
Define the linear operator $\Pi:\C^{2n}\oplus\C^{2n}\to\C^{2n}\oplus\C^{2n}$
as follows. Let $\Pi$ act on%

\[
\left(v_{-}\oplus v_{+}\right)\oplus\left(w_{-}\oplus w_{+}\right)\in\left(V_{-}\oplus V_{+}\right)\oplus\left(V_{-}\oplus V_{+}\right)=\C^{2n}\oplus\C^{2n},
\]
by
\[
\Pi:\left(v_{-}\oplus v_{+}\right)\oplus\left(w_{-}\oplus w_{+}\right)\mapsto\left(v_{-}\oplus w_{+}\right)\oplus\left(v_{+}\oplus w_{-}\right).
\]
For $P\in{\bf U}_{n,n}$ (or more general $P\subset\C^{2n}\oplus\C^{2n}$),
the \termi{Potapov transform} $\Pi(P)\subset\C^{2n}\oplus\C^{2n}$
of $P$ is defined by
\[
\Pi(P):=\left\{ \Pi x|x\in P\right\} .
\]

\begin{shaded}%
\begin{prop}
For any $P\in{\bf U}_{n,n}$, $\Pi(P)$ is a graph of an linear map
$\C^{2n}\to\C^{2n}$, and hence $\Pi(P)$ is identified with a $(2n\times2n)$
matrix.
\end{prop}

\end{shaded}

\begin{shaded}%
\begin{cor}
Let $P_{k}\in{\bf U}_{n,n}$ ($k\in\N$). Then $\lim_{k\to\infty}P_{k}$
converges in $G_{4n,2n}(\C)$ iff $\lim_{k\to\infty}\Pi(P_{k})$ converges
in $\Mat(2n,\C)$.
\end{cor}

\end{shaded}

For $g\in\Mat(2n,\C)$, let $\Pi(g):=\Pi(\graph(g))$.

\begin{shaded}%
\begin{prop}
{\rm{}(\cite{Ner11}: Sec.8.1 in Ch.2)} For $g=\left(\begin{matrix}a & b\\
c & d
\end{matrix}\right)\in\Mat(2n,\C)$ with $\det a\neq0$,
\begin{equation}
\Pi(g)=\left(\begin{matrix}-a^{-1}b & a^{-1}\\
d-ca^{-1}b & ca^{-1}
\end{matrix}\right).\label{eq:2.8.3}
\end{equation}
Note that if $g\in\Spn{\C}$, then $d-ca^{-1}b=a^{\TT-1}$.%
\end{prop}

\end{shaded}

\begin{example}
Let $X:={\footnotesize\begin{pmatrix}1 & 0\\
0 & -1
\end{pmatrix}}$ and consider $e^{tX}$ for $t\ge0$. Since $e^{tX}\in\Sp(2,\C)$
and $\cI_{2}-\left(e^{tX}\right)^{*}\cI_{2}e^{tX}\ge0$, we have $e^{tX}\in\Gamma\Sp_{c}(2)$.
We see
\[
\lim_{t\to+\infty}\Pi(e^{tX})=\lim_{t\to+\infty}\left(\begin{matrix}0 & e^{-t}\\
e^{-t} & 0
\end{matrix}\right)=O.
\]
Thus
\[
P:=\lim_{t\to+\infty}\graph(e^{tX})=\Pi^{-1}(\graph(O))=\left\{ (z,0,0,z')^{\TT}|z,z'\in\C\right\} \in\C^{4}\cong\C^{2}\oplus\C^{2},
\]
\[
\ker P=\left\{ (z,0)^{\TT}|z\in\C\right\} ,\qquad\indef P=\left\{ (0,z')^{\TT}|z'\in\C\right\} .
\]
Therefore we conclude that $P\in\bSp_{2}$, and that $\lim_{t\to+\infty}e^{tX}$
does not converge in $\Gamma\Sp_{c}(2)$, but does in $\bSp_{2}$.\qed
\end{example}

\begin{shaded}%
\begin{prop}
\label{thm:2.8.1}{\rm{}(\cite{Ner11}: Theorem 8.1, 8.2, 9.3 in
Ch.2)} We have
\begin{align}
\Pi(\rU(n,n)) & =\left\{ \left(\begin{matrix}\alpha & \beta\\
\gamma & \delta
\end{matrix}\right)\in\mathrm{U}(2n)\Big|\ \text{conditions }(1)\text{--}(4)\right\} ,\label{eq:Pi(U(n,n))=00003D}\\
\Pi(\Gamma\rU(n,n)) & =\left\{ r:=\left(\begin{matrix}\alpha & \beta\\
\gamma & \delta
\end{matrix}\right)\in\Mat(2n,\C)\Big|\ \|r\|\le1,\ \text{conditions }(1)\text{--}(4)\right\} ,\label{eq:Pi(GammaU)}\\
\Pi({\bf U}_{n,n}) & =\left\{ r:=\left(\begin{matrix}\alpha & \beta\\
\gamma & \delta
\end{matrix}\right)\in\Mat(2n,\C)\Big|\ \|r\|\le1,\ \text{conditions }(3),(4)\right\} ,\label{eq:Pi(Unn)}
\end{align}
where the conditions are (1) %
$\det\beta\neq0$, (2) %
$\det\gamma\neq0$, (3) $\Vert\alpha\Vert<1$, (4) $\Vert\delta\Vert<1$;
and the operator norm $\|r\|$ is w.r.t.\ the usual norm $\|v\|^{2}:=\left\langle v|v\right\rangle =v^{*}v$
on $\C^{2n}$. As to (\ref{eq:Pi(U(n,n))=00003D}), in fact, the conditions
(1)--(4) are equivalent, and so one of them suffices. Also note that
if $\|r\|\le1$, then $(1)\then(3)\&(4)$ and $(2)\then(3)\&(4)$.
\end{prop}

\end{shaded}%

\begin{shaded}%
\begin{prop}
\label{thm:Ner2.9.4}{\rm{}(\cite{Ner11}: Theorem 9.4 in Ch.2)}%
{} Let $P_{1},P_{2}\in{\bf U}_{n,n}$, and $\Pi(P_{1}),\Pi(P_{2})\in\Mat(2n,\C)$
be expressed by 
\[
\Pi(P_{1})=\left(\begin{matrix}\alpha & \beta\\
\gamma & \delta
\end{matrix}\right),\qquad\Pi(P_{2})=\left(\begin{matrix}\varphi & \psi\\
\theta & \varkappa
\end{matrix}\right).
\]
Then%
\begin{equation}
\Pi(P_{1}P_{2})=\left(\begin{matrix}\alpha+\beta(1-\varphi\delta)^{-1}\varphi\gamma & \beta(1-\varphi\delta)^{-1}\psi\\
\theta(1-\delta\varphi)^{-1}\gamma & \varkappa+\theta\delta(1-\varphi\delta)^{-1}\psi
\end{matrix}\right).\label{eq:2.8.9}
\end{equation}
\end{prop}

\end{shaded}

Let $\rho:\Gamma\Sp_{c}(2n)\to\Bdd(\cH)/\C^{\times}$ be the metaplectic
representation. 

\begin{shaded}%
\begin{thm}
\label{thm:Sp2n-rep}{\rm{}(\cite{Ner11}: Theorem 8.2, 8.5 in Ch.1,
Theorem 1.5 in Ch.5)} There exists a strongly continuous projective
representation $\rho':\bSp_{2n}\to\Bdd(\cH)/\C^{\times}$ which extends
$\rho$; That is, 
\[
\rho'(\graph(g))=\rho(g),\qquad g\in\Gamma\Sp_{c}(2n)
\]
\end{thm}

\end{shaded}

We call $\rho'$ the \termi{extended metaplectic representation},
and we write $\rho$ for $\rho'$ in the following. To obtain the
explicit formula for calculating $\rho(P)$ for any $P\in\bSp_{2n}$,
one may consider the Schr\"odinger representation on $L^{2}(\R^{2n})$
or the Fock--Bargmann representation on the boson Fock space $\cF(\C^{n})$.
In both cases $\rho(P)$ is expressed as an integral operator whose
kernel is a Gaussian function. (Precisely, in the former case, the
kernel can be a tempered distribution, not a function.) The Gaussian
kernel is written in terms of the {Potapov transform} of $P$ in
an explicit form.

\section{Results for ${\rm Sp}_{c}(2m,\protect\R)$}

\label{sec:Results-for}

Our first goal (\ref{eq:siborihA}) is achieved by Corollary \ref{cor:limrhoConverge}
and Theorem \ref{thm:slimexp=00003Dexp} below, which follows from
the following theorem. Recall $\cI_{2n}=(-I_{m})\oplus(-I_{m})\oplus I_{m}\oplus I_{m}$
and $\cN_{b}=O_{m}\oplus I_{m}\oplus O_{m}\oplus(-I_{m}).$

\begin{shaded}%
\begin{thm}
\label{thm:graphconverge}Let $n=2m$ and $\cA\in\spcN n{\R}$. %
Then
\[
\lim_{\nnn\to\infty}\graph\,e^{\cA-\nnn\cN_{b}}
\]
converges to some $P\in\bSp_{2n}$, explicitly written as follows.
Let $V^{(\lambda)}\subset\C^{4m}$ be the eigenspaces of $\cN_{b}$,
corresponding to the eigenvalues $\lambda=0,\pm1$ of $\cN_{b}$.
Then%
{} 
\begin{equation}
P=\left\{ \left(v_{-1}+v_{0}\right)\oplus\left(e^{\cA_{0}}v_{0}+v_{1}\right)|v_{\lambda}\in V^{(\lambda)},\ \lambda=0,\pm1\right\} ,\label{eq:P=00003D}
\end{equation}
where $\SYM{\cA_{0}}{A0}:=\left(I-\cN_{b}^{2}\right)\cA\left(I-\cN_{b}^{2}\right)$.
Hence we have%
\begin{equation}
\ker P=V^{(-1)}\subset\ker(\cI_{2n}-I),\qquad\indef P=V^{(1)}\subset\ker(\cI_{2n}+1),\label{eq:ker=00003Dindef=00003D}
\end{equation}
which verifies $P\in\bSp_{2n}$.
\end{thm}

\end{shaded}
\begin{proof}
There exists $\eps>0$ and smooth maps $(-\eps,\eps)\ni\epsilon\mapsto V_{\epsilon}^{(\lambda)}$
($\lambda=0,\pm1$) s.t. $V_{\epsilon}^{(\lambda)}$ is an invariant
subspace of $\epsilon\cA-\cN_{b}$ for each $\epsilon\in(-\eps,\eps)$
and $\lambda$, and that $V_{0}^{(\lambda)}=V^{(\lambda)}$. Let $P_{\epsilon}^{(\lambda)}$
be the corresponding projection onto $V_{\epsilon}^{(\lambda)}$;
Precisely, $P_{\epsilon}^{(\lambda)}$ is determined by
\[
\SYM{P_{\epsilon}^{(\lambda)}}{Pela}v:=\begin{cases}
v & \text{if}\quad v\in V_{\epsilon}^{(\lambda)}\\
0 & \text{if}\quad v\in V_{\epsilon}^{(\lambda')},\ \lambda'\neq\lambda
\end{cases}.
\]
 Let
\[
\SYM{A_{\epsilon}^{(\lambda)}}{Aela}:=\left(\epsilon\cA-\cN_{b}\right)P_{\epsilon}^{(\lambda)},
\]
Set $\epsilon=1/\nnn$. We have 
\[
\graph\,e^{\cA-\nnn\cN_{b}}=\graph\,e^{\nnn\left(\epsilon\cA-\cN_{b}\right)}=\graph\,\exp\left(\nnn\sum_{\lambda}A_{\epsilon}^{(\lambda)}\right)=\bigoplus_{\lambda}\graph\left(\exp\left(\nnn A_{\epsilon}^{(\lambda)}|_{V_{\epsilon}^{(\lambda)}}\right)\right).
\]
Since $\lim_{\nnn\to\infty}A_{\epsilon}^{(\lambda)}=A_{0}^{(\lambda)}=\lambda P_{0}^{(\lambda)}$,
we have
\begin{align*}
\lim_{\nnn\to\infty}\exp\nnn A_{\epsilon}^{(-1)} & =P_{0}^{(0)}+P_{0}^{(1)},\\
\lim_{\nnn\to\infty}\exp\left(-\nnn A_{\epsilon}^{(1)}\right) & =P_{0}^{(-1)}+P_{0}^{(0)},\\
\lim_{\nnn\to\infty}\exp\nnn A_{\epsilon}^{(0)} & =\exp\dot{A}_{0}^{(0)},
\end{align*}
where $\dot{A}_{0}^{(0)}:=\frac{d}{d\epsilon}A_{\epsilon}^{(0)}\Big|_{\epsilon=0}$.
Thus the following hold:
\begin{align*}
\lim_{\nnn\to\infty}\graph\exp\left(\nnn A_{\epsilon}^{(-1)}|_{V_{\epsilon}^{(-1)}}\right) & =\left\{ v\oplus\left(P_{0}^{(0)}+P_{0}^{(1)}\right)v|v\in V^{(-1)}\right\} =\left\{ v\oplus0|v\in V^{(-1)}\right\} ,\\
\lim_{\nnn\to\infty}\graph\exp\left(\nnn A_{\epsilon}^{(1)}|_{V_{\epsilon}^{(1)}}\right) & =\left\{ \left(P_{0}^{(-1)}+P_{0}^{(0)}\right)v\oplus v|v\in V^{(1)}\right\} =\left\{ 0\oplus v|v\in V^{(1)}\right\} ,\\
\lim_{\nnn\to\infty}\graph\exp\left(\nnn A_{\epsilon}^{(0)}|_{V_{\epsilon}^{(0)}}\right) & =\left\{ v\oplus\left(\exp\dot{A}_{0}^{(0)}\right)v|v\in V^{(0)}\right\} .
\end{align*}
Therefore we find that $\graph\left(e^{\cA-\nnn\cN_{b}}\right)$ converges
as $\nnn\to\infty$, and that
\begin{align}
\lim_{\nnn\to\infty}\graph\left(e^{\cA-\nnn\cN_{b}}\right) & =\left\{ v\oplus0|v\in V^{(-1)}\right\} \oplus\left\{ v\oplus\left(\exp\dot{A}_{0}^{(0)}\right)v|v\in V^{(0)}\right\} \oplus\left\{ 0\oplus v|v\in V^{(1)}\right\} \\
 & =\left\{ \left(v_{-1}+v_{0}\right)\oplus\left(\left(\exp\dot{A}_{0}^{(0)}\right)v_{0}+v_{1}\right)|v_{\lambda}\in V^{(\lambda)},\ \lambda=0,\pm1\right\} .\label{eq:limgraph1}
\end{align}
Thus by Lemma \ref{thm:d/dePe0=00003D} below, we find (\ref{eq:P=00003D})
and (\ref{eq:ker=00003Dindef=00003D}). %
\end{proof}
\begin{shaded}%
\begin{lem}
\label{thm:d/dePe0=00003D}%
\textup{
\begin{equation}
\frac{d}{d\epsilon}\Big|_{\epsilon=0}P_{\epsilon}^{(0)}=\cN_{b}\cA\left(I-\cN_{b}^{2}\right)+\left(I-\cN_{b}^{2}\right)\cA\cN_{b},\label{eq:dPs0}
\end{equation}
}and
\begin{equation}
\dot{A}_{0}^{(0)}=\cA_{0}:=\left(I-\cN_{b}^{2}\right)\cA\left(I-\cN_{b}^{2}\right).\label{eq:ddeA0e}
\end{equation}
\end{lem}

\end{shaded}
\begin{proof}

To prove (\ref{eq:dPs0}), observe that $P_{\epsilon}^{(0)}$ is explicitly
written as
\[
P_{\epsilon}^{(0)}=\lim_{j\to\infty}P_{\epsilon,j}^{(0)},\qquad\SYM{P_{\epsilon,j}^{(0)}}{P0ej}:=\left[\left(2\left(\epsilon\cA-\cN_{b}\right)\right)^{2j}+I\right]^{-1},
\]
and that if $t\mapsto X_{t}$ is any matrix-valued smooth function
$\R\to{\rm GL}(n,\C)$, then
\begin{equation}
\frac{d}{dt}\Big|_{t=0}X_{t}^{-1}=-X_{0}^{-1}\left(\frac{d}{dt}\Big|_{t=0}X_{t}\right)X_{0}^{-1}.\label{eq:ddtXt}
\end{equation}
Since
\begin{equation}
P_{0,j}^{(0)}=\left(\left(2\cN_{b}\right)^{2j}+I\right)^{-1}=\left(I-\cN_{b}^{2}\right)+\left(2^{2j}+1\right)^{-1}\cN_{b}^{2}=I+\left[-1+\left(2^{2j}+1\right)^{-1}\right]\cN_{b}^{2}\label{eq:P00n}
\end{equation}
and%
\begin{equation}
\frac{d}{d\epsilon}\Big|_{\epsilon=0}\left(P_{\epsilon,j}^{(0)}\right)^{-1}=-2^{2j}\left[\cA\cN_{b}+\cN_{b}\cA+(j-1)\cN_{b}^{2}\cA\cN_{b}+(j-1)\cN_{b}\cA\cN_{b}^{2}\right],\label{eq:ddeP}
\end{equation}
we find from (\ref{eq:ddtXt}) that
\[
\frac{d}{d\epsilon}\Big|_{\epsilon=0}P_{\epsilon,j}^{(0)}=-P_{0,j}^{(0)}\left[\frac{d}{d\epsilon}\Big|_{\epsilon=0}\left(P_{\epsilon,j}^{(0)}\right)^{-1}\right]P_{0,j}^{(0)}.
\]
Substituting (\ref{eq:P00n}) and (\ref{eq:ddeP}), a straightforward
calculation shows that%
\begin{align*}
 & \frac{d}{d\epsilon}\Big|_{\epsilon=0}P_{\epsilon,j}^{(0)}=-\Big[-\frac{2^{2j}}{2^{2j}+1}\cN_{b}\cA\left(I-\cN_{b}^{2}\right)+2^{2j}\left(2^{2j}+1\right)^{-1}\\
 & \qquad\times\left\{ -\left(I-\cN_{b}^{2}\right)\cA\cN_{b}-\frac{1}{2^{2j}+1}\cN_{b}^{2}\left[\cA\cN_{b}+\cN_{b}\cA+(j-1)\cN_{b}^{2}\cA\cN_{b}+(j-1)\cN_{b}\cA\cN_{b}^{2}\right]\cN_{b}^{2}\right\} \Big].
\end{align*}
Therefore we find%
\[
\lim_{j\to\infty}\frac{d}{d\epsilon}\Big|_{\epsilon=0}P_{\epsilon,j}^{(0)}=\cN_{b}\cA\left(I-\cN_{b}^{2}\right)+\left(I-\cN_{b}^{2}\right)\cA\cN_{b}
\]
and hence (\ref{eq:dPs0}) holds. %
We can verify (\ref{eq:ddeP}) by%
\begin{align*}
\dot{A}_{0}^{(0)} & =\frac{d}{d\epsilon}A_{\epsilon}^{(0)}\Big|_{\epsilon=0}=\frac{d}{d\epsilon}\left(\epsilon\cA-\cN_{b}\right)P_{\epsilon}^{(0)}\Big|_{\epsilon=0}=\cA P_{0}^{(0)}-\cN_{b}\frac{d}{d\epsilon}P_{\epsilon}^{(0)}\Big|_{\epsilon=0}\ \\
 & =\cA P_{0}^{(0)}-\cN_{b}\left[\cN_{b}\cA\left(I-\cN_{b}^{2}\right)+\left(I-\cN_{b}^{2}\right)\cA\cN_{b}\right]\qquad(\text{by (\ref{eq:dPs0})})\\
 & =\cA(I-\cN_{b}^{2})-\cN_{b}^{2}\cA\left(I-\cN_{b}^{2}\right)-\cN_{b}\left(I-\cN_{b}^{2}\right)\cA\cN_{b}\\
 & =\left(I-\cN_{b}^{2}\right)\cA\left(I-\cN_{b}^{2}\right)=\cA_{0}.
\end{align*}

\end{proof}
\begin{shaded}%
\begin{cor}
\label{cor:limrhoConverge} $\slim_{\nu\to\infty}\rho(e^{t(\cA-\nnn\cN_{b})})$
converges to some $\C^{\times}T_{\cA,t}\in\Bdd(\cH)/\C^{\times}$
for each $t>0$, where
\end{cor}

\begin{enumerate}
\item $T_{\cA,t}\in\Bdd(\cH)$ is strongly continuous w.r.t.\ $t>0$, 
\item $\|T_{\cA,t}\|=1$ for all $t>0$,
\item $\slim_{t\searrow0}T_{\cA,t}=E_{b}$. %
{} 
\end{enumerate}
\end{shaded}
\begin{proof}
Clear from Theorem \ref{thm:Sp2n-rep} and \ref{thm:graphconverge}. 
\end{proof}

The following theorem is a special case of Conjecture \ref{conj:sibori}.

\begin{shaded}%
\begin{thm}
\label{thm:slimexp=00003Dexp}Let $\cA\in\spcN m{\R}$. Then we have
\begin{align}
\slim_{\nnn\to\infty}\exp\bigl(h_{\cA}(\vec{Z})-\nnn N_{b}\bigr) & =\exp\bigl[\cE_{b}(h_{\cA}(\vec{Z}))\bigr]E_{b}.\label{eq:slimexp=00003Dexp}
\end{align}
\end{thm}

\end{shaded}
\begin{proof}
By Lemma \ref{thm:hA=00003Ddrho}, we find
\begin{align}
\slim_{\nnn\to\infty}\C^{\times}\exp\left(h_{\cA}(\vec{Z})-\nnn N_{b}\right) & =\slim_{\nnn\to\infty}\rho\left(\exp\left(\hat{\cA}-\nnn\cN_{b}\right)\right).\label{eq:slimslim}
\end{align}
 Hence by Corollary \ref{cor:limrhoConverge}, %

\begin{equation}
\slim_{\nnn\to\infty}\zeta_{t,\nnn}\exp t\left(h_{\cA}(\vec{Z})-\nnn N_{b}\right)=T_{\cA,t}\in\Bdd(\cH),\label{eq:slimzeta}
\end{equation}
for some sequence $\zeta_{t,1},\zeta_{t,2},....\in\C^{\times}$, for
each $t>0$. We can assume that $\zeta_{t,\nnn}$ depends on $t$
as $\zeta_{t,\nnn}=e^{t\xi_{\nnn}}$, so that 
\[
T_{\cA,t}=\slim_{\nnn\to\infty}T_{\cA,t}^{(\nnn)},\quad T_{\cA,t}^{(\nnn)}:=\exp t\left(h_{\cA}(\vec{Z})-\nnn N_{b}+\xi_{\nnn}\right)\quad\text{ and }\quad T_{\cA,s}T_{\cA,t}=T_{\cA,s+t}.
\]
Let $T_{\cA,0}:=\slim_{t\searrow0}T_{\cA,t}=E_{b}$. Then $\left\{ T_{\cA,t}|t\ge0\right\} $
is a strongly continuous one-parameter semigroup, but this is not
a contraction semigroup in the usual sense (e.g.\ \cite[X.8]{RS75}),
since $T_{\cA,0}\neq I$. %
{} Noticing 
\[
T_{\cA,t}\psi=\lim_{\epsilon\searrow0}T_{\cA,t+\epsilon}\psi=\lim_{\epsilon\searrow0}T_{\cA,\epsilon}T_{\cA,t}\psi=E_{b}T_{\cA,t}\psi\in\cK:=\ran\,E_{b}=\ker N_{b},\qquad\psi\in\cH,
\]
instead we consider $\left\{ T_{\cA,t}\!\upha_{\cK}|t\ge0\right\} $,
which is a strongly continuous contraction semigroup of bounded operators
on $\cK$ in the usual sense%
. Hence there exists a closed operator $X_{\cA}$ densely defined
on $\cK$ s.t.\ $T_{\cA,t}\upha_{\cK}$ is formally expressed as
$e^{tX_{\cA}}$. Let $\psi\in\cK\cap\dom(h_{\cA}(\vec{Z}))$, then%
\[
\lim_{\nnn\to\infty}E_{b}\frac{d}{dt}T_{\cA,t}^{(\nnn)}\psi\Big|_{t=0}=\lim_{\nnn\to\infty}E_{b}\left(h_{\cA}(\vec{Z})-\nnn N_{b}+\xi_{\nnn}\right)\psi=E_{b}h_{\cA}(\vec{Z})\psi+\lim_{\nnn\to\infty}\xi_{\nnn}\psi.
\]
Hence the above l.h.s.\ converges iff $\lim_{\nnn\to\infty}\xi_{\nnn}$
converges. If we set $\xi_{\nnn}\equiv0$ ($\zeta_{t,\nnn}\equiv1$),
then we have 
\[
X_{\cA}\psi=\frac{d}{dt}T_{\cA,t}\psi\Big|_{t=0}=\lim_{\nnn\to\infty}E_{b}\frac{d}{dt}T_{\cA,t}^{(\nnn)}\psi\Big|_{t=0}=E_{b}h_{\cA}(\vec{Z})\psi.
\]
Eq.\,(\ref{eq:slimexp=00003Dexp}) follows from this.
\end{proof}

\begin{lyxgreyedout}
\hidec{
\begin{proof}
\[
\cG:=\lim_{\nu\to\infty}\graph\exp\left(i\nu R+H\right)=\lim_{\nu\to\infty}\graph\exp\nu\left(iR+\epsilon H\right)=\lim_{\nu\to\infty}\graph\exp\nu\left(A_{\epsilon}^{(-1)}+A_{\epsilon}^{(0)}+A_{\epsilon}^{(1)}\right)=
\]
\[
=_{???}\lim_{\nu\to\infty}\graph\exp\nu\left(A_{\epsilon}^{(-1)}|_{V_{\epsilon}^{(-1)}}\right)\oplus\graph\exp\nu\left(A_{\epsilon}^{(0)}|_{V_{\epsilon}^{(0)}}\right)\oplus\graph\exp\nu\left(A_{\epsilon}^{(1)}|_{V_{\epsilon}^{(1)}}\right)
\]
\[
=_{???}\left\{ v\oplus0|v\in V^{(-1)}\right\} \oplus\left\{ v\oplus\left(\exp\dot{A}_{0}^{(0)}\right)v|v\in V^{(0)}\right\} \oplus\left\{ 0\oplus v|v\in V^{(1)}\right\} 
\]
\[
\ker\cG=V^{(-1)}
\]
\[
\indef\cG=V^{(1)}
\]
\end{proof}
Since $\cG$ is $\Lambda^{\ominus}$-Lagrangian, and $V^{(0)}$ is
a symplectic subspace of $\C^{4n}$ 
\[
\exp\dot{A}_{0}^{(0)}\in\Sp(V^{(0)})
\]
and so $I\oplus\exp\dot{A}_{0}^{(0)}\oplus I\in\Sp(4n,\C)$

\rule[0.5ex]{1\columnwidth}{1pt}

\[
\lim_{\nu\to\infty}\graph\exp i\nu R=\left\{ v\oplus0|v\in V^{(-1)}\right\} \oplus\left\{ v\oplus v|v\in V^{(0)}\right\} \oplus\left\{ 0\oplus v|v\in V^{(1)}\right\} 
\]
Since

\[
R\dot{P}_{0}^{(0)}=\lim_{\epsilon\to0}\epsilon^{-1}RP_{\epsilon}^{(0)}
\]

\[
r:=iR\qquad I,\quad r^{k}Hr^{l},\quad0\le k,l\le2
\]

\[
\dot{A}_{0}^{(0)}=cI+\sum_{k,l=0}^{2}c_{kl}r^{k}Hr^{l}
\]

\[
J:=\begin{pmatrix} & I\\
-I
\end{pmatrix}\qquad X:=\left(\begin{matrix}\alpha & \beta\\
\gamma & -\alpha^{t}
\end{matrix}\right)\in\sp(2n,\C)
\]

\begin{lyxgreyedout}
recall: The Lie algebra $\sp(2n,\bbK)$, of the group $\Sp(2n,\bbK)$,
consists of real matrices

\[
X=\left(\begin{matrix}\alpha & \beta\\
\gamma & \delta
\end{matrix}\right),\text{ where }\delta=-\alpha^{t},\ \beta=\beta^{t},\ \gamma=\gamma^{t}
\]
\[
\iff\cJ X+X^{\TT}\cJ=0\qquad\cJ=\begin{pmatrix}0 & I\\
-I & 0
\end{pmatrix}
\]
\begin{lyxgreyedout}
\[
\cJ X=\begin{pmatrix}0 & I\\
-I & 0
\end{pmatrix}\left(\begin{matrix}\alpha & \beta\\
\gamma & \delta
\end{matrix}\right)=\begin{pmatrix}\gamma & \delta\\
-\alpha & -\beta
\end{pmatrix}\qquad X^{\TT}\cJ=\begin{pmatrix}\alpha^{\TT} & \gamma^{\TT}\\
\beta^{\TT} & \delta^{\TT}
\end{pmatrix}\begin{pmatrix}0 & I\\
-I & 0
\end{pmatrix}=\begin{pmatrix}-\gamma^{\TT} & \alpha^{\TT}\\
-\delta^{\TT} & \beta^{\TT}
\end{pmatrix}
\]
\end{lyxgreyedout}
\end{lyxgreyedout}
\begin{equation}
\mathfrak{W}\left(\begin{matrix}\alpha & \beta\\
\gamma & -\alpha^{t}
\end{matrix}\right):=-\frac{1}{2}\tr\alpha-\frac{1}{2}\sum_{k,l}\beta_{kl}z_{k}z_{l}-\sum_{k,l}\alpha_{kl}z_{k}\frac{\partial}{\partial z_{l}}+\frac{1}{2}\sum_{k,l}\gamma_{kl}\frac{\partial^{2}}{\partial z_{k}\partial z_{l}}.\qquad(4.2)\label{eq:6.4.2-1}
\end{equation}
\[
=-\frac{1}{2}\tr\alpha-\frac{1}{2}\sum_{k,l}\beta_{kl}a_{k}^{*}a_{l}^{*}-\sum_{k,l}\alpha_{kl}a_{k}^{*}a_{l}+\frac{1}{2}\sum_{k,l}\gamma_{kl}a_{k}a_{l}
\]
\[
=\frac{1}{2}\vec{A}^{\TT}JX\vec{A}\qquad\vec{A}=\begin{pmatrix}\vec{a}\\
\vec{a}^{*}
\end{pmatrix}
\]

\begin{proof}
\begin{lyxgreyedout}
\[
\vec{A}^{\TT}JX\vec{A}=(\vec{a}^{\TT},\vec{a}^{*\TT})\begin{pmatrix} & I\\
-I
\end{pmatrix}\left(\begin{matrix}\alpha & \beta\\
\gamma & -\alpha^{t}
\end{matrix}\right)\begin{pmatrix}\vec{a}\\
\vec{a}^{*}
\end{pmatrix}=(\vec{a}^{\TT},\vec{a}^{*\TT})\begin{pmatrix} & I\\
-I
\end{pmatrix}\begin{pmatrix}\alpha\vec{a}+\beta\vec{a}^{*}\\
\gamma\vec{a}-\alpha^{\TT}\vec{a}^{*}
\end{pmatrix}=
\]
\[
=(\vec{a}^{\TT},\vec{a}^{*\TT})\begin{pmatrix}\gamma\vec{a}-\alpha^{\TT}\vec{a}^{*}\\
-\left(\alpha\vec{a}+\beta\vec{a}^{*}\right)
\end{pmatrix}=\vec{a}^{\TT}\left(\gamma\vec{a}-\alpha^{\TT}\vec{a}^{*}\right)-\vec{a}^{*\TT}\left(\alpha\vec{a}+\beta\vec{a}^{*}\right)
\]
\[
=\vec{a}^{\TT}\gamma\vec{a}-\vec{a}^{\TT}\alpha^{\TT}\vec{a}^{*}-\vec{a}^{*\TT}\alpha\vec{a}-\vec{a}^{*\TT}\beta\vec{a}^{*}
\]
\[
=\sum_{kl}a_{k}\gamma_{kl}a_{l}-\sum_{kl}a_{k}\alpha_{lk}a_{l}^{*}-\sum_{kl}a_{k}^{*}\alpha_{kl}a_{l}-\sum_{kl}a_{k}^{*}\beta_{kl}a_{l}^{*}
\]
\[
=\sum_{kl}a_{k}\gamma_{kl}a_{l}-\sum_{kl}\alpha_{lk}\left(a_{l}^{*}a_{k}+\delta_{kl}\right)-\sum_{kl}a_{k}^{*}\alpha_{kl}a_{l}-\sum_{kl}a_{k}^{*}\beta_{kl}a_{l}^{*}
\]
\[
=-\tr\alpha-\sum_{kl}a_{k}^{*}\beta_{kl}a_{l}^{*}-2\sum_{kl}a_{k}^{*}\alpha_{kl}a_{l}+\sum_{kl}a_{k}\gamma_{kl}a_{l}
\]
\end{lyxgreyedout}
\end{proof}

}%
\end{lyxgreyedout}

\section{Differential operators}

\label{sec:diffop}

{} Let $m\in\N$, $n:=2m$. Consider the representations of annihilation/creation
operators $a_{k},b_{k},a_{k}^{*},b_{k}^{*}$ ($k=1,...,m$) on $L^{2}(\C^{m})$
where $Z_{k}=a_{k}^{*}+b_{k}$ is represented by $z_{k}$ (the $k$th
coordinate of $\C^{m}$, viewed as a multiplication operator on $L^{2}(\C^{m})$)
for all $k$. An example of such representation is given by 
\[
a_{k}:=\frac{\di}{\di z_{k}}+\frac{\ol z_{k}}{2},\qquad b_{k}:=\frac{\di}{\di\ol z_{k}}+\frac{z_{k}}{2},
\]
and so
\[
a_{k}^{*}=-\frac{\di}{\di\ol z_{k}}+\frac{z_{k}}{2},\qquad b_{k}^{*}=-\frac{\di}{\di z_{k}}+\frac{\ol z_{k}}{2}.
\]
Let
\[
\Re b_{k}:=\frac{1}{2}\left(b_{k}+b_{k}^{*}\right)=\frac{1}{2}\left(\im\frac{\di}{\di y_{k}}+x_{k}\right),\qquad\Im b_{k}:=\frac{1}{2\im}\left(b_{k}-b_{k}^{*}\right)=\frac{1}{2}\left(-\im\frac{\di}{\di x_{k}}+y_{k}\right),
\]
where $z_{k}=x_{k}+\im y_{k}$, $x_{k},y_{k}\in\R$. Then we have
the following:

\[
\left[\Re b_{k},\Im b_{k}\right]=\frac{\im}{2},\qquad b_{k}^{*}b_{k}=(\Re b_{k})^{2}+(\Im b_{k})^{2}-\frac{1}{2},
\]
and hence
\begin{align*}
N_{b} & =\sum_{k=1}^{m}\left[(\Re b_{k})^{2}+(\Im b_{k})^{2}\right]-\frac{m}{2}\\
 & =\frac{1}{4}\sum_{k=1}^{m}\left[\left(\im\frac{\di}{\di y_{k}}+x_{k}\right)^{2}+\left(-\im\frac{\di}{\di x_{k}}+y_{k}\right)^{2}\right]-\frac{m}{2}.
\end{align*}
Let $\alpha=(\alpha_{1},...,\alpha_{2m}):\R^{2m}\to\R^{2m}$ be a
smooth map, where $\alpha_{k}:\R^{2m}\to\R\ (k=1,...,2m)$. Then $\alpha$
is seen as a smooth vector field on $\R^{2m}\cong\C^{m}$. The (positive-definite)
\termi{magnetic Laplacian} $\Delta^{\alpha}$ is defined by 
\[
\Delta^{\alpha}:=-\sum_{k=1}^{2m}\left(\frac{\di}{\di x_{k}}+\im\alpha_{k}\right)^{2},
\]
where we set $x_{m+k}:=y_{k}$ for $k=1,...,m$. (For more general,
geometric and rigorous definitions of magnetic Laplacians, see e.g.\ \cite{Shu01}.)
Set
\[
\alpha_{k}(\vec{x}):=x_{m+k},\qquad\alpha_{m+k}(\vec{x}):=-x_{k},\qquad k=1,...,m,\ \vec{x}:=(x_{1},...,x_{2m}),
\]
then we have
\[
N_{b}=\frac{1}{4}\Delta^{\alpha}-\frac{m}{2}.
\]

In this representation, the operator $h_{\cA}(\vec{Z})$ is nothing
other than $h_{\cA}$ as a multiplication operator on $L^{2}(\C^{m})$.
Thus by Theorem \ref{thm:slimexp=00003Dexp}, we have the following:

\begin{shaded}%
\begin{cor}
\label{thm:slimexphDelta}For any $\cA\in\spcN m{\R}$, 
\begin{align}
\slim_{\nnn\to\infty}\exp\left(h_{\cA}-\nnn\left(\frac{1}{4}\Delta^{\alpha}-\frac{m}{2}\right)\right) & =\exp\left[\cE_{b}(h_{\cA})\right]E_{b}.\label{eq:slimexphDelta}
\end{align}
\end{cor}

\end{shaded}

\section{Path integral representation}

\label{sec:pathint}Let $\cH:=L^{2}(\C^{m})$ and $\cK:=\ran\,E_{b}=\ker N_{b}\subset\cH$.
Let $\Omega_{0}\in\cK$ be a unit vector such that $N_{a}\Omega_{0}=0$,
which is unique up to scalar multiples; equivalently, $a_{k}\Omega_{0}=0$
for all $k=1,...,m$. We call it the \termi{vacuum vector}. We also
fix a unit vector $\Omega_{\vec{z}}\in\cK$ for each $\vec{z}=(z_{1},...,z_{m})\in\C^{m}$
such that
\[
a_{k}\Omega_{\vec{z}}=z_{k}\Omega_{\vec{z}},\quad k=1,...,m.
\]
These are called \termi{coherent vectors}.
\[
\p_{\vec{z}}:=\ket{\Omega_{\vec{z}}}\bra{\Omega_{\vec{z}}}
\]
\[
\d\mu(\vec{z}):=\frac{1}{\pi^{m}}\d x_{1}\d y_{1}\cdots\d x_{m}\d y_{m},\qquad z_{k}=x_{k}+\im y_{k},
\]
\[
\int_{\C^{m}}\p_{\vec{z}}\d\mu(\vec{z})=E_{b}\text{ on }L^{2}(\C^{m}).
\]
We find that the manifold $\Manifold:=\left\{ \p_{\vec{z}}|\vec{z}\in\C^{m}\right\} \cong\C^{m}$,
where each $\p_{\vec{z}}$ is viewed as an operator on $\cK$, not
on $L^{2}(\C^{m})$, is an example of more general notion of the \termi{family of coherent states}
defined in \cite{Yam22a}. Then we find that the antinormally ordered
quantization $\cE_{b}(f)$ of a function $f:\C^{m}\to\R$ is expressed
as
\[
\cE_{b}(f)=\cQ(f):=\int_{\C^{m}}f(\vec{z})\p_{\vec{z}}\d\mu(\vec{z}).
\]
From Sec.~7 of \cite{Yam22a}, we know that Corollary 5.8 in \cite{Yam22a}
holds for this case. This implies the following.

For $T>0$, let $\mu_{[0,T]}$ be the standard Brownian bridge measure
on the path space
\[
\SYM{{\rm loop}_{[0,T],0}}{loop[]}:=\left\{ \varphi\in C([0,T],\R^{2m})|\varphi(0)=\varphi(T)=0\right\} .
\]
For $\nnn>0$ and $\varphi\in{\rm loop}_{[0,\nnn],0}$, define $\varphi^{(\nnn)}\in\SYM{{\rm loop}_{0}}{loop0}:={\rm loop}_{[0,1],0}$
by $\SYM{\varphi^{(\nnn)}(t)}{phit(no)}:=\varphi(\nnn t)$. 

The measure $\mu_{[0,\nnn]}$ and the bijection ${\rm loop}_{[0,\nnn],0}\to{\rm loop}_{0}$,
$\varphi\mapsto\varphi^{(\nnn)}$, induce the measure $\mu_{\nnn}$
on ${\rm loop}_{0}$. Let

\[
\alpha^{\flat}:=\sum_{k=1}^{m}\left(y_{k}\d x_{k}-x_{k}\d y_{k}\right),\qquad z_{k}=x_{k}+\im y_{k}.
\]
This is denoted by $\theta_{{\rm nor}}$ in Sec.~7 of \cite{Yam22a}.%
\begin{shaded}%
\begin{thm}
\label{thm:vacextPathInt-bdd}If $H:\C^{m}\to\R$ is smooth and bounded,
, 
\begin{align}
\bigl\langle\Omega_{0}|e^{-\im\cE_{b}(H)}\Omega_{0}\bigr\rangle=\lim_{\nnn\to\infty}e^{\nnn m}\int_{{\rm loop}_{0}}e^{\im S_{H}(\varphi)}\d\mu_{\nnn}(\varphi)=\lim_{\nnn\to\infty}e^{\nnn m}\Ex_{\nnn}\Bigl[e^{\im S_{H}(\varphi)}\Bigr]\label{eq:vacexp=00003Dexp-bdd}
\end{align}
where
\[
S_{H}(\varphi):=\int_{\varphi}\alpha^{\flat}+\int_{0}^{1}H(\varphi(t))\d t.
\]
Here, the line integral $\int_{\varphi}\alpha^{\flat}$ of the 1-form
$\alpha^{\flat}$ along the curve $\varphi$ is understood as a stochastic
integral in the sense of Stratonovich w.r.t.\ the path measure $\mu_{\nnn}$.
$\Ex_{\nnn}[\cdot]$ denotes the \emph{(classical)} expectation value
w.r.t.~the probability measure $\mu_{\nnn}$.
\end{thm}

\end{shaded}

Recall that $h_{\cA}$ is defined by (\ref{eq:def:hA}), and pure
imaginary. For nonzero $\cA\in\sp(2m,\R)$, we see $h_{\cA}$ is unbounded.
Hence we cannot set $H=\im h_{\cA}$ in the above theorem.%
{} Instead we take a simple workaround: For $\no>0$, define the cutoff
Hamiltonian $h_{\cA}^{(\no)}:\C^{m}\to\R$ by
\[
h_{\cA}^{(\no)}(\vec{z}):=\frac{1}{\im}\max\{\min\{\im h_{\cA}(\vec{z}),\no\},-\no\}.
\]
This is bounded but not smooth. Although the above theorem assumes
the smoothness of $H$, it is not used in its proof in \cite{Yam22a};
We assumed it simply because it usually holds for classical mechanical
systems. Hence we can set $H=\im h_{\cA}^{(\no)}$ in the above theorem. 

Here we recall the following three theorems:

\begin{shaded}%
\begin{thm}
\label{thm:Trotter}{\rm{}\cite[Theorem VIII.21]{RS75}%
} Let $\{A_{n}\}$ and $A$ be self-adjoint operators. Then $A_{n}\rightarrow A$
in the strong resolvent sense if and only if $e^{iA_{n}}$ converges
strongly to $e^{itA}$ for each $t.$
\end{thm}

\end{shaded}

\begin{shaded}%
\begin{thm}
\label{thm:commonCore}{\rm{}\cite[Theorem VIII.25]{RS75}} Let
$\{A_{n}\}_{n=1}^{\infty}$ and $A$ be self-adjoint operators and
suppose that $D$ is a common core for all $A_{n}$, $A$; in other
words, that they are essentially self-adjoint on $D$. If $A_{n}\varphi\rightarrow A\varphi$
for each $\varphi\in D,$ then $A_{n}\to A$ in the strong resolvent
sense.
\end{thm}

\end{shaded}

\begin{shaded}%
\begin{thm}
\label{thm:Nelson}{\rm{}\cite[Theorem X.39]{RS80} (Nelson's analytic
vector theorem)} Let $A$ be a symmetric operator on a Hilbert space
$\cH$. If $\dom(A)$ contains a total set of analytic vectors, then
$A$ is essentially self-adjoint.
\end{thm}

\end{shaded}

\begin{shaded}%
\begin{prop}
For each $\cA\in\spcN m{\R}$ and $t\in\R$, we have
\begin{equation}
\slim_{\no\to\infty}e^{t\cE_{b}(h_{\cA}^{(\no)})}=e^{t\cE_{b}(h_{\cA})}.\label{eq:slimeiteit}
\end{equation}
\end{prop}

\end{shaded}
\begin{proof}
We know that the vectors in the dense subspace $D$ of ``finite particle
states'' in $\cK$ w.r.t.~the creation/annihilation operators $a_{k}^{*},a_{k}$
are analytic vectors of $\im\cE_{b}(h_{\cA})$; See e.g.~\cite[p.190]{Fol89}.
(Alternatively we could take the space of finite linear combinations
of coherent vectors.) Hence $\im\cE_{b}(h_{\cA})$ is essentially
self-adjoint on $D$ by Theorem \ref{thm:Nelson}. ($\im\cE_{b}(h_{\cA}^{(\no)})$
is also essentially self-adjoint on $D$ since it is bounded.)%
{} For any $v\in D$, we easily find that
\[
\lim_{\no\to\infty}\cE_{b}(h_{\cA}^{(\no)})v=\lim_{\no\to\infty}E_{b}h_{\cA}^{(\no)}v=E_{b}h_{\cA}v=\cE_{b}(h_{\cA})v.
\]
Thus by Theorem \ref{thm:Trotter} and \ref{thm:commonCore}, we have
(\ref{eq:slimeiteit}).
\end{proof}
\begin{shaded}%
\begin{cor}
\label{thm:vacextPathInt}Let $\cA\in\spcN m{\R}$.%
{} With the notations of Theorem \ref{thm:vacextPathInt-bdd}, we have
\begin{align}
\bigl\langle\Omega_{0}|e^{-\cE_{b}(h_{\cA})}\Omega_{0}\bigr\rangle=\lim_{\no\to\infty}\lim_{\nnn\to\infty}e^{\nnn m}\Ex_{\nnn}\Bigl[e^{\im S_{\cA,\no}(\varphi)}\Bigr]\label{eq:vacexp=00003Dexp-unbdd}
\end{align}
where
\[
S_{\cA,\no}(\varphi):=\int_{\varphi}\alpha^{\flat}+\int_{0}^{1}\im h_{\cA}^{(\no)}(\varphi(t))\d t\in\R.
\]
Here, the line integral $\int_{\varphi}\alpha^{\flat}$ is understood
as a stochastic integral in the sense of Stratonovich w.r.t.\ the
path measure $\mu_{\nnn}$. 
\end{cor}

\end{shaded}
\begin{acknowledgement*}
I am grateful to Koichi Arashi for for useful discussions on unitary
representations of non-compact Lie groups.
\end{acknowledgement*}
{} %

% ref. http://rexpit.blog29.fc2.com/blog-entry-124.html
%\clearpage
\phantomsection	% ← hyperref を使う時は入れる。使わないなら不要。 	
\addcontentsline{toc}{section}{Reference}

% \bibliographystyle{plain}
% \bibliography{ybib}

%<index label>

%<index label dense>
\providecommand{\noopsort}[1]{}\providecommand{\singleletter}[1]{#1}%

\end{document}